# Synthesis of Single-crystal-like Nanoporous Carbon Membranes and Their Application in Overall Water Splitting

Hong Wang[1], Shixiong Min[2], Chun Ma[1], Zhixiong Liu[1], Weiyi Zhang[3], Qiang Wang[4], Debao Li[4], Yangyang Li[1], Stuart Turner[5], Yu Han[6], Haibo Zhu[2], Edy Abou-hamad[2], Mohamed Nejib Hedhili[7], Jun Pan[1], Weili Yu[1], Kuo-Wei Huang[2], Lain-Jong Li[1], Jiayin Yuan[3]*, Markus Antonietti[3], Tom Wu[1]*

[1]Physical Science and Engineering Division, King Abdullah University of Science & Technology, Thuwal, 23955-6900, Saudi Arabia.

[2]Kaust Catalysis Center, King Abdullah University of Science & Technology, Thuwal, 23955-6900, Saudi Arabia.

[3]Max Planck Institute of Colloids and Interfaces, Department of Colloid Chemistry, D-14476 Potsdam, Germany.

[4]State Key Laboratory of Coal Conversion, Institute of Coal Chemistry, The Chinese Academy of Sciences, Taiyuan 030001 People's Republic of China

[5]EMAT, University of Antwerp, Groenenborgerlaan 171, B-2020 Antwerpen, Belgium.

[6]Membrane Research Centre, King Abdullah University of Science and Technology, Thuwal 23955-6900, Saudi Arabia.

[7]Imaging and Characterization Core Lab, King Abdullah University of Science and Technology, Thuwal, 23955-6900 Saudi Arabia.

*Corresponding authors: Jiayin.Yuan@mpikg.mpg.de; tao.wu@kaust.edu.sa;

**Nanoporous graphitic carbon membranes with defined chemical composition and pore architecture are novel nanomaterials that are actively pursued. Compared to easy-to-make porous carbon powders that dominate the porous carbon research and applications in energy generation/conversion and environmental remediation, porous carbon membranes are synthetically more challenging though rather appealing from an application perspective due to their structural integrity, interconnectivity and purity. Here we report a simple bottom-up approach to fabricate large-size, freestanding, porous carbon membranes that feature an unusual single-crystal-like graphitic order and hierarchical pore architecture plus favorable nitrogen doping. When loaded with cobalt nanoparticles, such carbon membranes serve as high-performance carbon-based non-noble metal electrocatalyst for overall water splitting.**

Carbon materials have been widely used to address global energy and environmental issues due to their extraordinary, tuneable physicochemical properties, rich abundance and low cost[1-3]. Freestanding porous carbon membranes particularly hold great promise in the fields of catalysis, water treatment, biofiltration, gas separation and optoelectronics, just to name a few, due to their structural integrity, continuity, and purity[4-5]. Typical synthetic methods involve mechanical rolling of thermally expanded graphite flakes, chemical vapour deposition and vacuum filtration of dispersions of graphene sheets or carbon nanotubes[6-10]. In addition, Koros and co-works reported that pyrolysis of thermosetting polymer precursors (*e.g.* aromatic polyimides) could lead to carbon membrane sieves with micropores, which exhibited high-performance for gas separation[11-13]. For some carbon-based energy applications, such as electrodes in electrochemical energy conversion/storage, and nanoelectronic devices, however, precise control over the atomic order, local chemical composition, nanoscale morphology and complex pore architecture, as well as easy access to porous membranes of large size and large surface area, is highly relevant but cannot be fully met by the state-of-the-art synthetic protocols. Particularly, a high degree of graphitization and hierarchical pore architecture with interconnected pores over a broad length scale are eagerly being pursued because they could offer fast electron conduction, and rapid mass transport through large pores along with a simultaneously high reaction capacity *via* the large accessible surface area provided by the micro/mesopores[14]. Furthermore, the pores in the cross sections of the carbon membrane if distributed in a gradient manner, can offer unconventional fluidic transport on the nanoscale (*e.g.*, concentration gradient[15] and permselectivity[16, 17] for broad application in micro/nano-fluidic devices[18, 19]).

In this study, we reported a bottom-up approach for fabrication of hierarchically structured, nitrogen-doped, graphitic nanoporous carbon membranes (termed HNDCMs) *via* morphology-retaining carbonization of a porous polymer membrane precursor. In particular, the pores along the membrane cross section assumed a gradient distribution in their sizes, and the pore walls exhibited unusual single-crystal-like characteristics. As a prototypical application, when loaded with cobalt nanoparticles, these highly conductive porous carbon membranes served as an active carbon-based bifunctional electrocatalyst for overall water splitting.

**Results**

**Synthesis and structure characterizations.** Fig. 1a shows the membrane fabrication process. First, porous polyelectrolyte membranes bearing gradually varying pore sizes along the membrane cross section (termed GPPMs) were assembled according to a previously reported procedure by exploiting electrostatic crosslinking, *i.e.* interpolyelectrolyte complexation between cationic poly[1-cyanomethyl-3-vinylimidazolium bis(trifluoromethanesulfonyl)imide] (PCMVImTf$_2$N) and anionic neutralized poly(acrylic acid) (PAA)[20]. The structural characterization of PCMVImTf$_2$N (Supplementary Figs. 1-3) and details of the polymer membrane fabrication method are provided in the supporting materials. It is important to note that the preparation of GPPMs is a mature and robust technique that can produce various pore profiles at desirable size scales. Next, direct pyrolysis of the freestanding GPPMs under a nitrogen flow yielded HNDCMs. For example, carbonization of a rectangular GPPM that was 7.2 x 3.3 cm$^2$ in size and 96 μm in thickness (Fig. 1b) produced a HNDCM that was 5.2 x 2.5 cm$^2$ in size and 62 μm thick (Fig. 1c). Shrinkage of the membrane dimensions during pyrolysis was accompanied by a weight reduction of 75%.

Importantly, the pore architectures of the carbon membranes can be regulated by the molecular weight (MW) of the polymeric precursors. This correlation was investigated by pairing the same PCMVImTf$_2$N with PAAs of different MW. Here, the GPPM-x and HNDCM-x-y notations are used, where x and y denote the MW of PAA and the carbonization temperature, respectively. These two crucial parameters are carefully paired to prepare carbon membranes with the desirable characteristics. For example, GPPM-2000 exhibited an interconnected porous network but its carbon product at 1000 °C (*i.e.*, HNDCM-2000-1000) only possessed inconsecutive pores (Supplementary Fig. 4). In fact, the interconnected pores in GPPM-2000 were blocked even at 300 °C (Supplementary Fig. 5). Surprisingly, pyrolysis of GPPM-100,000 (Supplementary Fig. 6) at 1000 °C preserves the well-defined porous morphology (Fig. 1d), and an asymmetric, three-dimensionally interconnected pore architecture was spontaneously created in HNDCM-100,000-1000. From the top to the bottom, the pore size of HNDCM-100,000-1000 gradually decreased from 1.5 μm and 900 nm to 550 nm in zones I, II and III. Impressively, in sample HNDCM-250,000-1000, the pore sizes (Figs. 1e-h) decreased to 250 ±10 nm, 75 ± 8 nm, and 32 ± 6 nm in zones I, II and III, respectively, indicating that the pore size can be readily tuned by the MW of PAA. As observed, the pore morphology of the carbon membrane is position-specific, *i.e.* from larger ones on the top gradually to smaller ones at the bottom. It is actually a natural outcome of the crosslinking density profile in the GPPM template. The higher the crosslinking density in the polymer membrane is, the larger the pores in the carbon membrane will be, because dense crosslinking undermines the trimerization reaction of cyano groups that require sufficient mobility to reach desirable positions to complete the triazine ring formation. Therefore, the spatial restriction promotes thermodegradation of the porous polymer network in the highly crosslinked top region, which is further amplified by larger pores in the same area, forming carbon pores with larger size than those at the bottom. For PAA with an even higher MW of 450,000 and 3,000,000, the carbon membranes were highly porous but became too fragile upon carbonization (Supplementary Figs. 7, 8). In general, pyrolysis enlarges the pore size in HNDCMs compared to that in GPPMs due to the considerable mass loss in the form of volatile species during carbonization (Supplementary Fig. 9).

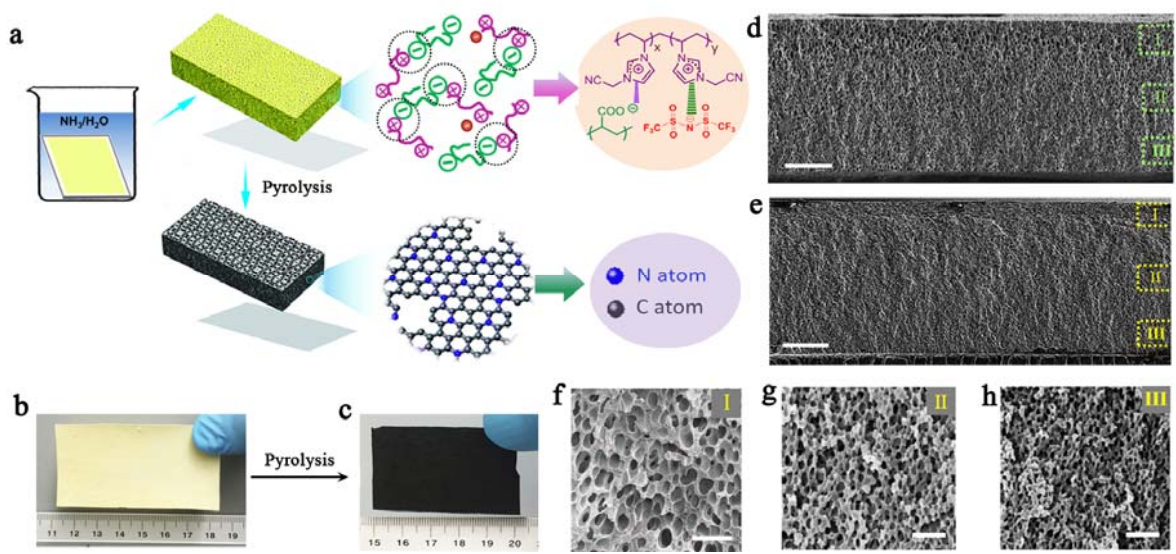

**Figure 1 | Formation and structure of hierarchically structured nitrogen-doped porous carbon membranes.** (**a**) Schematic illustration of the preparation procedure. (**b**) Photograph of a 7.2 x 3.3 cm$^2$ freestanding GPPM. (**c**) Photograph of a 5.2 x 2.5 cm$^2$ freestanding HNDCM obtained by pyrolysis of GPPM in **b,** the unit in the ruler in (**b**) and (**c**) is centimeter. (**d**) Cross-section scanning electron microscopy (SEM) image of the HNDCM-100,000-1000. (**e**) SEM image of the cross section of HNDCM-250,000-1000, the scale bars in (**d**) and (**e**) are 20 μm. (**f-h**) High-magnification SEM images of the cross-section structures of HNDCM-250,000-1000. The scale bars represent 500 nm.

**Formation mechanism.** The relationship between the porous morphology and the MW of PAA provides a practical route to tailor the membrane structure, which is a natural outcome of the function of PAA in the synthesis of GPPMs. PAA acts as a crosslinker to chemically lock PCMVImTf$_2$N in a porous network *via* electrostatic complexation. In addition, the crosslinking density in the GPPMs increased as the MW of PAA increased (Supplementary Table 1). The collapse of GPPM-2000 at temperatures above 300 °C was due to the relatively low crosslinking density (*i.e.*, pores too large), which cannot stabilize the pores; the cracking of the carbon membranes prepared from PAA with a MW of ~ 450,000 and 3,000,000 resulted from a crosslinking density that was too high (*i.e.*, pores too small), which results in the build-up of excessively high inner stress during the carbonization process. Only pyrolysis of polymer membranes with moderate MW ends up with retention of the structural integrity. According to the thermogravimetric analysis (Supplementary Fig. 10), these polymer membranes start to detach their alkyl chains and H and O elements at 280 °C, and a thermally induced trimerization reaction of the cyano groups simultaneously occurs to build up a stable triazine network[21]. It is important to note that in general, morphology retaining carbonization *via* pyrolysis of porous polymer precursors is challenging because pyrolysis typically breaks down polymeric chains and results in cracks[22]. Dai *et al.* reported that the cross-linking state of polymer precursors is a key to achieving crack-free carbon membranes[23]. Here, our synthesis of HNDCMs demonstrates that the porous nano/microstructure in the carbon precursor can be preserved due to a synergistic combination of the initial crosslinked state of the precursors and the subsequent formation of a thermally stable network intermediate during the bottom-up carbonization process.

**Unique graphitic structures.** High-resolution transmission electron microscopy (HRTEM) images provide insight into the microscopic and atomic structures of the HNDCM-100,000-y samples (y=800, 900, and 1000) that were prepared using three different pyrolysis temperatures. Fig. 2a shows the presence of mesopores that are 2~50 nm in size for HNDCM-100,000-800. Interestingly, as shown in Figs. 2b, 2c and 2d, we observed onion-like concentric graphitic nanostructures that consist of multi-shells and hollow cage-like centres. The shells are composed of (002) graphitic planes with a lattice spacing of $0.338 \pm 0.02$ nm. The HRTEM images of HNDCM-100,000-800 displayed in Figs. 2e and Supplementary Fig. 11 show the preferential orientation of the graphitic layers. Unexpectedly, in HNDCM-100,000-900, a single-crystal-like atomic packing emerged in the entire membrane. The fringes show a well-defined lattice spacing of $0.196 \pm 0.02$ nm (Figs. 2f and Supplementary Fig.12), which corresponds to the (101) plane of graphite[24]. Notably, HNDCM-100,000-1000 has the same graphitic structure but with fewer lattice defects (Figs. 2g and Supplementary Fig. 13). A selected area electron diffraction measurement (inset in Fig. 2g) yielded a six-fold symmetric spot pattern, which is characteristic of well-crystalized graphite. These results indicate progressive graphitization at elevated temperatures from 800 to 1000 °C. A similar trend was confirmed by Raman, XRD and solid-state $^{13}$C-NMR measurements (Supplementary Figs. 14-16). Energy-filtered transmission electron microscopy mappings for both C and N (Fig. 2h) indicate a uniform distribution of N in the carbon matrix, which is expected due to *in situ* molecular doping of HNDCM with N. The synergy between the N lone pair and the π-system of the C lattice can dramatically alter the physicochemical properties of the HNDCMs (*e.g.*, oxidative stability and catalytic activity)[25]. For example, the HNDCM-100,000-1000 sample is fire-retardant (Supplementary Fig. 17). Even in an acetylene flame (> 1000 °C) in air for 60 s, this sample maintained its original colour and morphology, which is indicative of its excellent oxidative stability and its potential for use as a fire-retardant protective material.

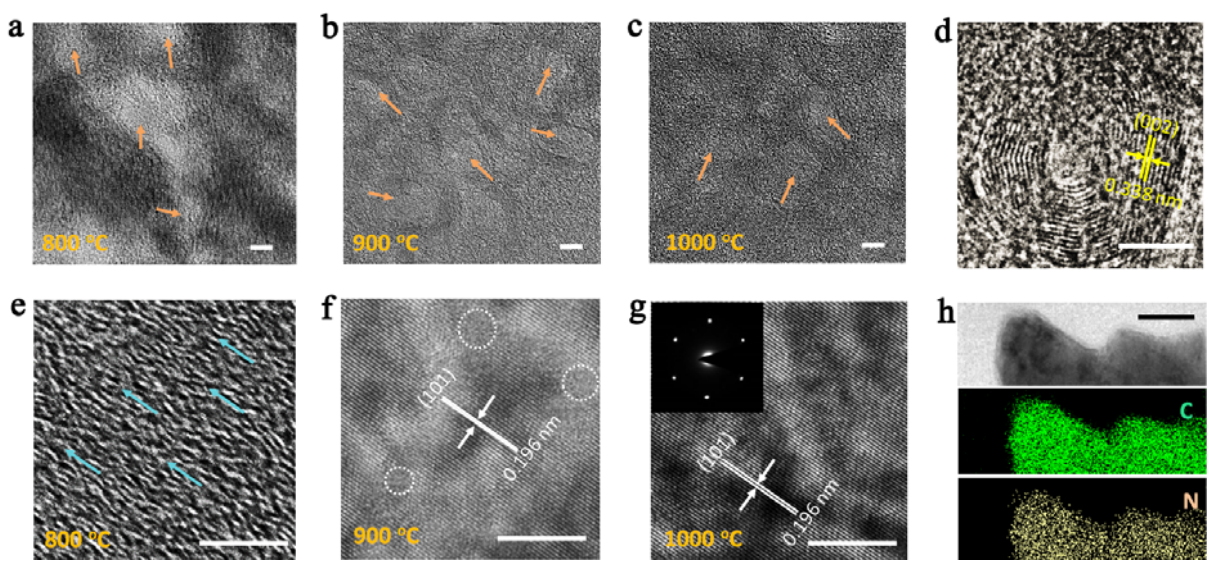

**Figure 2 | Microstructural characterizations of the N-doped porous carbon membrane.** HRTEM images of (**a, e**) HNDCM-100,000-800, (**b, f**) HNDCM-100,000-900, (**c, g**) HNDCM-100,000-1000, and (**d**) a typical onion-like graphitic structure in HNDCM-100,000-1000. Some defective regions are highlighted in **f**. Inset in (**g**) is the SAED pattern for HNDCM-100,000-1000. The SAED pattern indicates the single-crystal-like characteristics of HNDCM-100,000-1000. The scale bars of (**a-g**) are 5 nm; (**h**) TEM image and corresponding elemental (C and N) mappings, scale bar : 50 nm.

The elemental analysis indicated that the N content in HNDCM-100,000-800/900/1000 membranes was 11.7 wt.%, 8.27 wt. %, and 5.7 wt.%, respectively, which is in good accordance with the results from the X-ray photoelectron spectroscopy (XPS) analysis. As previously reported, a high N content may hinder the crystallinity of carbon[26], which is in good agreement with our observation that despite a relatively lower N content, HNDCM-100,000-1000 is more graphitic than the other two samples. Nevertheless, the high crystallinity of the membranes prepared at 900 and 1000 °C is surprising because the formation of graphitic carbon from polymer precursors typically requires much higher pyrolysis temperatures. Previous studies have demonstrated that pyrolysis of carbon precursors above 800 °C in the presence of metal catalysts can improve graphitization[27]. However, HNDCMs are free of metal catalyst, as extra confirmed by XPS measurements (Supplementary Fig. 18). Single-crystal-like carbons cannot be obtained by carbonization at 1000 °C of either native PCMVImTf$_2$N or its physical blend with PAA (Supplementary Fig. 19) as well as other polymeric precursors, such as polyacrylonitrile[28] and poly(acrylamide-*co*-acrylic acid)[29]. It is important to note that PCMVImTf$_2$N rather than PAA is the main carbon precursor for HNDCMs due to its high carbonization yield and being 75 wt% of the GPPMs. Furthermore, the poorly porous carbon membrane of HNDCM-2,000-1000 is dominantly amorphous with graphitic domains that only surround the pores (Supplementary Fig. 20). This result implies that the graphitization may be facilitated by the highly porous precursors at temperatures lower than those required for nonporous or poorly porous ones, which is a phenomenon that most likely arises from the abundant high-energy surfaces in the porous structures. The graphite sheets are not stable upon size shrinkage and tend to rearrange into concentric graphitic shells due to the beneficial effect of symmetric and uniform strain distribution[30]. Furthermore, higher temperatures reduced the diameter of the hollow cages and introduced more shells in the onion-like structures of HNDCMs (Supplementary Figs. 21, 22), which is consistent with previous reports[30, 31]. Based on these analyses, the observed unique graphitic order in the HNDCM-100,000-900/1000 samples resulted from the porous precursor that facilitated migration and recrystallization of the carbon atoms into graphite. In fact, our results suggest that the graphitization started at temperature as low as 900 °C. Due to the small and/or thin size of the macropore wall, these graphite sheets in the carbon membrane preferentially rearrange themselves into the lowest energetically state.

**Specific surface area and conductivity characterizations.** Fig. 3a shows the $N_2$ absorption-desorption isotherms of HNDCM-100,000-y (y=800, 900 and 1000), and the results indicate that the pore volume and specific surface area ($S_{BET}$) significantly increased as the temperature increased. The $S_{BET}$ of HNDCM-100,000-800/-900/-1000 was 354, 632, and 907 $m^2\ g^{-1}$, respectively, and their total pore volumes were 0.48, 0.61, and 0.79 $cm^3\ g^{-1}$, respectively. The sharp increase of $S_{BET}$ at low pressures ($P/P_0 < 0.05$) is due to the nitrogen filling in micropores below 2 nm, which is confirmed by the density functional theory (DFT) pore size distribution curves (Fig. 3b) derived from the $N_2$ adsorption branches. The obvious hysteresis above $P/P_0 \sim 0.5$ is indicative of the existence of mesopores. In previous studies, $Tf_2N^-$ in the polymer matrix was reported to act as the micropore forming agent[32]. In our case, $Tf_2N^-$ constitutes 53.8 wt % of GPPM-100,000 based on elemental analysis (Supplementary Fig. 23), and is thereby responsible for the formation of micropores and small mesopores. In these porous membranes with hierarchical architectures, micropores and small mesopores are beneficial and provide active surface areas with high accessibility, and the large mesopores and macropores form interconnected three-dimensional networks and serve as transport highways to accelerate mass diffusion and significantly promote exchange efficiency.

An additional advantage of the highly crystalline graphite structure is the high conductivity of the carbon membranes despite their high pore volume. For example, the conductivity of HNDCM-100,000-1000 reached its highest value (*i.e.*, 200 S cm$^{-1}$) at 298 K, which decreased to 147 and 32 S cm$^{-1}$ at 298 K for HNDCM-100,000-900 and HNDCM-100,000-800, respectively. This high conductivity is appealing for a wide range of electrical/electrochemical applications. Furthermore, the conductivity of HNDCM-100,000-y (y=800, 900 and 1000) increased with the test temperature, which is characteristic of semiconductor-like behaviour (Fig. 3c). Notably, the conductivity of HNDCM-100,000-1000 is one of the highest values ever reported for macroscopic carbon monoliths (Fig. 3d) (ref. 33-36).

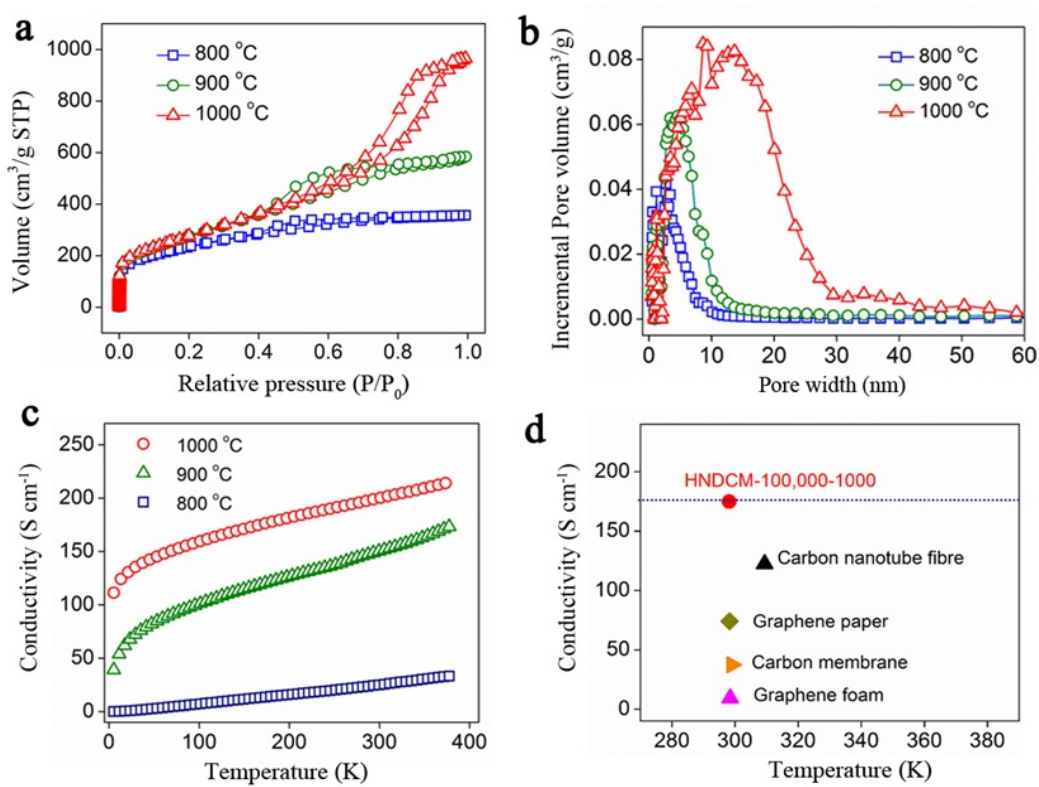

**Figure 3 | Brunauer-Emmett-Teller (BET) specific surface area and conductivity characterizations.** (a) N$_2$ absorption-desorption isotherms and (b) corresponding pore size distribution of HNDCM-100,000-1000/900/800. (c) Temperature dependence of the conductivity measured for HNDCM-100,000-1000/900/800 from 5 K to 390 K using a four-probe method. (d) Comparison of the conductivity of HNDCM-100,000-1000 with previous results for macroscopic carbon materials (*e.g.*, carbon nanotube fibre[33], graphene paper[34], carbon membrane[35] and graphene foam[36]).

**Functionalization and electrochemical performance.** Polyelectrolyte-derived complexes can bind and immobilize metal ions, salts, and nanoparticles[37], which inspires us to explore the functionalization of HNDCMs with metal nanoparticles *via* doping the polymeric precursors with metal species. Recently, the conversion of renewable energy resources *via* water splitting to $H_2$ and $O_2$ is of primary urgency to address issues associated with global warming and energy crisis[38]. Scalable and sustainable electrochemical water splitting is a promising technology. However, this approach requires highly efficient, robust earth-abundant electrocatalyst materials to replace the costly Pt catalyst[39]. To date, remarkable hydrogen evolution reaction (HER) and oxygen evolution reaction (OER) electrocatalysts have been applied in water splitting[40, 41]. Owing to their thermodynamic convenience and potential applications in proton-exchange membranes or alkaline electrolysers, most efforts in this field have been devoted to developing HER and OER catalysts that function in strongly acidic and basic conditions, respectively.[42-44] However, to accomplish overall water splitting, the coupling of HER and OER catalysts in the same electrolyte is desirable from the viewpoint of simplification of the system and cost reduction.[45-53] Here, the HNDCM-100,000-1000 sample bearing embedded cobalt nanoparticles (termed HNDDC-100,000-1000/Co) was investigated as an active bifunctional electrocatalyst for overall water splitting in alkaline media. HNDCM-100,000-1000/Co was chosen as an example due to its favourable high conductivity and large surface area. Details of the synthesis and structural characterizations are provided in the methods and supporting Information (Supplementary Figs. 24-26).

The SEM images (Figs. 4a-b) of HNDCM-100,000-1000/Co suggest that the pore architectures of the HNDCM are preserved during the carbonization in the presence of cobalt acetate. As shown in Fig. 4c and 4d, after treatment with 1 M aqueous hydrochloric acid (HCl) solution for 12 hs, two types of Co nanoparticles were found uniformly distributed throughout the carbon membrane. One type is the ultrafine Co nanoparticles with a diameter of 1.2 ± 0.5 nm (Figure 4c), and the other is larger Co nanoparticles with a diameter of 20 ± 2 nm covered by a thin graphitic carbon shell of several nm in thickness (Figure 4d and Supplementary Fig.26). Previous reports demonstrated that downsizing heterogeneous catalysts in the nanoparticulate form (1-20 nm) could expose more active sites[54]. In addition, it was reported that excess of Co content in N-doped graphene could decrease its HER activity[55]. In accordance with these reports, an etching treatment which removes excessive Co nanoparticles was found to be very beneficial to improve the electrochemical activity and stability, as shown in the case of HNDCM-100,000-1000/Co.

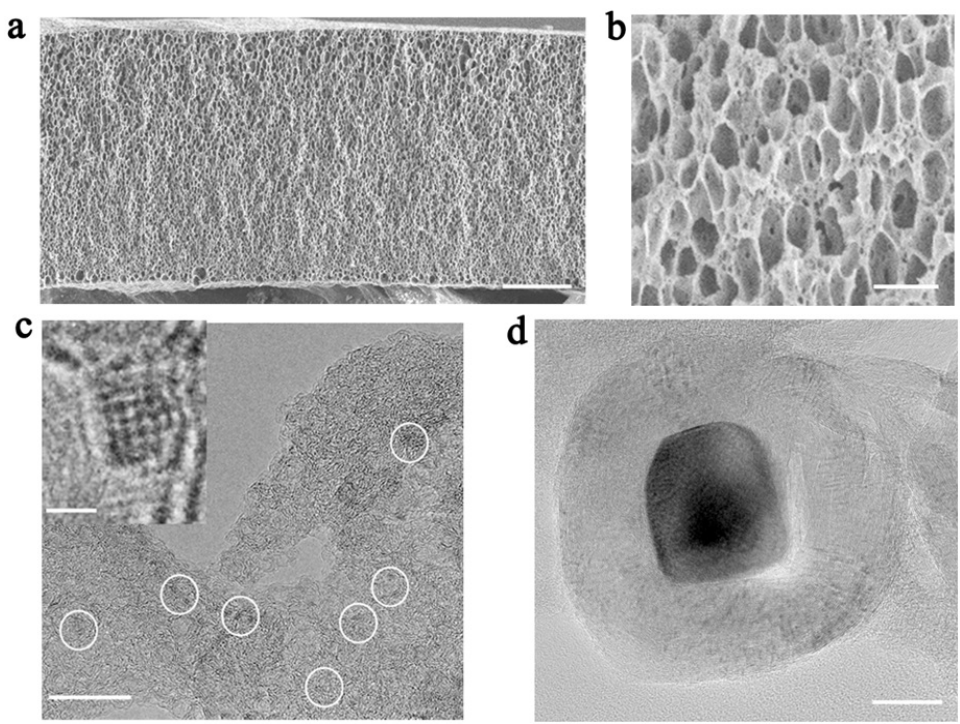

**Figure 4 | Microstructures of HNDCM-100,000-1000/Co.** (**a**) Cross-section SEM image, scale bar: 20 μm, (**b**) High-magnification SEM image, scale bar: 2 μm, and (**c-d**) HRTEM images of HNDCM-100,000-1000/Co obtained from different areas, scale bar: 10 nm; inset in (c) is the HRTEM image of a single ultrafine Co nanoparticle, scale bar: 1 nm.

The electrocatalytic performance of HNDCM-100,000-1000/Co was evaluated in 1 M KOH solution for both HER and OER. Fig. 5a shows the polarization curves obtained from linear sweep voltammetry (LSV) measurements, and a slow sweep rate of 1 mV s$^{-1}$ was used to eliminate any capacitance effect (see experimental detail in supporting information). HNDCM-100,000-1000/Co exhibited a high HER activity with a current density of 10 mA cm$^{-2}$ at an overpotential of 158 mV after IR-correction (the LSV data without IR-correction are provided in Supplementary Fig.27), which is significantly lower than that previously reported for Co/N-doped carbon nanotube catalyst[56]. In addition, this result is comparable or even superior to many other non-noble metal catalysts (Supplementary Table 2). The Tafel slope extracted from the LSV curve was determined to be 93.4 mV dec$^{-1}$ (Fig. 5c), indicating that the HER driven by this catalyst was controlled by a Volmer-Heyrovsky mechanism[57]. Fig. 5b shows the LSV curve for the OER. Here, a low overpotential of 199 mV was required to reach a current density of 10 mA cm$^{-2}$, and the Tafel slope was as small as 66.8 mV dec$^{-1}$ (Fig. 5d), close to the ideal value of 59 mV dec$^{-1}$ (equivalent to 2.3RT/F) associated with a one-electron transfer prior to the rate-limiting step[58]. These values outperform previous results on the Co or CoO$_x$/carbon hybrid OER catalysts[59, 60]. It is important to note that the Co loading in our catalyst was 2.16 wt% (determined by inductively coupled plasma-atomic emission spectroscopy), and therefore, the high catalytic activity of our catalyst for HER and OER is believed to result from a synergy between its high conductivity, nitrogen-doping, hierarchical pore architecture and high dispersion state of the active Co nanoparticles in HNDCM-100,000-1000. Most importantly, unlike previously reported HER electrocatalysts[61], no performance degradation was induced by bubble trapping in our catalyst due to the rapid mass transfer throughout the hierarchical pore architectures as well as the bubble-repelling surfaces of the nanostructures[62]. The "noise" in LSV curves for HER and OER was generated by perturbations in our membrane catalyst due to the release of large amounts of $H_2$ and $O_2$ bubbles that were produced at higher overpotentials. In general, for powder catalysts, polymer binders are used to process the catalyst films onto conductive substrates, which was avoided in our binder-free carbon membrane. Moreover, vigorous gas production can typically delaminate the active materials from the electrodes due to weakening of the binder, resulting in instability during their long-term operation. In contrast, our freestanding membrane catalyst is free of any polymer binder, leading to high stability for HER and OER (Fig. 5e). Meanwhile, the cyclic voltammetry (CV) durability tests of the 100,000-1000/Co electrodes for HER and OER were carried out (Figure S28). Compared with the long-term stability test of HNDCM-100,000-1000/Co electrode for HER and OER at a constant voltage, the CV stability data show that the 100,000-1000/Co electrodes can suffer slightly from a long-time reduction/oxidation process (Supplementary Fig.28 a). In the OER CV durability test data, there is a detectable decrease with time (Supplementary Fig28 b), which could be attributed to the irreversible reactions occurring at high potentials.

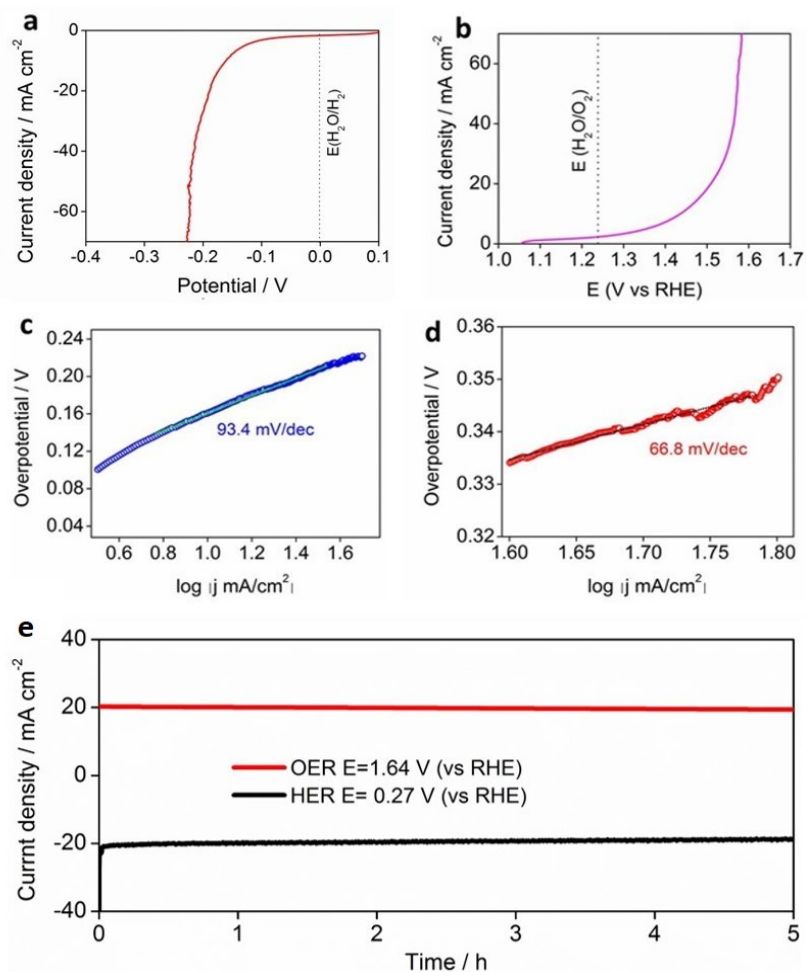

**Figure 5 | Electrocatalytic performance of HNDCM-100,000-1000/Co for overall water splitting in 1 M KOH.** (**a, b**) *J–V* curves after IR correction for HER and OER, respectively; (**c, d**) Tafel plots for the data presented in (a) and (b), respectively; (**e**) Long-term stability test result of the HNDCM-100,000-1000/Co electrode for HER and OER at a constant voltage.

**Discussion**

Our experimental results present a viable route toward preparing freestanding, nanoporous carbon membranes that feature an unusual single-crystal-like graphitic order and hierarchical pore architecture as well as favourable nitrogen doping. It was found that polymer precursors of moderate MW and the crosslinking state of polyelectrolyte membranes are crucial to achieve morphology-maintaining carbonization. Owing to the ionic character of the polyelectrolyte membranes and their nature of absorption and immobilization of metal ions, carbon membranes loaded with Co nanoparticles can be readily prepared through carbonization of polyelectrolyte membranes that absorbed cobalt ions. These hybrids are highly active non-noble metal electrocatalyst for overall water splitting.

Importantly, the synthesis and engineering of our membrane-like catalyst can be easily scaled up in size and quantity. As a proof-of-concept demonstration for solar-driven electrolysis, we used a commercially available 20 W solar panel to perform the HER on a piece of HNDCM-100,000-1000/Co film that was as large as 10.5 x 3.5 cm$^2$ (Supplementary Fig. 29), which is the maximum size limited by our carbonization oven. At a non-regulated output voltage of 20 V, an actual H$_2$ production rate of ~16 mL/min was achieved (Supplementary Fig. 30). This result indicates that our low-cost catalyst meets the requirements for industrial H$_2$ production in a large, clean manner in alkaline media. It is important to note that the OER is not only essential for water splitting but is also relevant for the charging process of rechargeable metal-air batteries[63]. The excellent OER activity in combination with the devisable shapes of our membrane-like catalyst affords a new avenue for the development of other efficient energy conversion devices. Furthermore, we expect that the electrocatalytic properties of the HNDCM-based hybrids can be further optimized by choosing appropriate metal species, and also thickness of the carbon membrane. The gradient pore architecture of the porous carbon membrane could offer an ideal platform for exploiting potential applications, such as electro-assisted separation and alternative energy conversion schemes.

**Methods**

**Materials and reagents** 1-Vinylimidazole (Aldrich 99%), 2,2'-azobis(2-methylpropionitrile) (AIBN, 98%), bromoacetonitrile (Aldrich 97%), and bis(trifluoromethane sulfonyl)imide lithium salt (Aldrich 99%) were used as received without further purifications. Dimethyl sulfoxide (DMSO), dimethyl formamide (DMF), methanol, and tetrahydrofuran (THF) were of analytic grade. Several poly(acrylic acid) samples (PAA) (MW: 2000 g/mol, solid powder; MW: 100,000 g/mol, 35 wt% in water, MW: 250,000 g/mol, 35 wt% in water; MW: 450,000 g/mol, solid powder; MW: 3,000,000 g/mol, solid powder ) were obtained from Sigma Aldrich. Poly[1-cyanomethyl-3-vinylimidazolium bis(trifluoromethanesulfonyl)imide] (PCMVImTf$_2$N) was prepared according to a previous report[64].

**Fabrication of the hierarchically structured porous nitrogen-doped carbon membranes.** First, the as-prepared gradient porous polymer membranes were clapped between two clean quartz plates and dried at 60 °C overnight under atmospheric pressure. For the carbonization process, the GPPMs were heated to 300 °C at a heating rate of 3 °C min$^{-1}$ under nitrogen flow, and held at 300 °C for one hour. They were then heated to the targeted carbonization temperature at a heating rate of 3 °C min$^{-1}$ under nitrogen flow. After holding at the final temperature for 1 h, the samples were cooled down to room temperature. During the process of carbonization, the vacuum in furnace was kept constant at 1.5 torr.

**Fabrication of Co-loaded HNDCM-100,000-1000 hybrid catalyst.** Freshly prepared GPPM-100,000 was placed in 200 mL of cobalt acetate aqueous solution (2 wt %) at pH ~ 5 adjusted with 0.1 M acetic acid. The mixture was refluxed at 80 °C for 24 h. Afterward, GPPM-100,000-Co(CH$_3$COO)$_2$ was taken out from the solution, washed with water, and dried at room temperature till constant weight. Finally, pyrolysis of GPPM-100,000-Co(CH$_3$COO)$_2$ was carried out similarly to that of the HNDCMs, leading to HNDCM-100,000-1000/Co. Before electrochemical test, the HNDCM-100,000-1000/Co membranes were immersed in 1M HCl solution for 12 hs to remove exposed Co nanoparticles sitting on the eternal surfaces of the porous carbon membranes.

**Electrochemical measurements.** The electrochemical measurements were performed in a conventional three electrode electrochemical cell using a CHI750E station. A graphite rod and an Ag/AgCl (in saturated KCl solution) electrode were used as the counter and reference electrodes, respectively. In order to avoid the possible effect of Pt deposition on our electrocatalyst during long-term electrochemical reaction, we selected graphite rod as counter electrode in all of the electrochemical reactions. The working electrode was fabricated by wrapping HNDCM-100,000-1000/Co free-standing film from one side with high-conductive copper tape, which is connected to a copper wire, and the exposed area of copper wires was covered with hot-melt glue to avoid the direct contact with electrolyte. The electrode area was calculated from its surface area. All the applied potentials are reported as reversible hydrogen electrode (RHE) potential scale using E (*vs*. RHE) = E (*vs*. Ag/AgCl) + 0.217 V + 0.0591 V*pH after IR correction. Potentiostatic EIS was used to determine the uncompensated solution resistance (Rs). The HER and OER activity of HNDCM-100,000-1000/Co treated with 1 M HCl was evaluated by measuring polarization curves with linear sweep voltammetry (LSV) technique at a scan rate of 1 mV/s in 1.0 M KOH (pH 14) solution. The stability tests for the HNDCM-100,000-1000/Co catalysts were performed using chronoamperometry at a constant applied overpotential.

**Characterization.** $^1$H- and solid state $^{13}$C-NMR spectra were recorded on a Bruker AVANCE III spectrometer operating at 400 and 100 MHz resonance frequencies, respectively. NMR chemical shifts were measured with respect to tetramethylsilane (TMS) as an external reference. X-ray diffraction (XRD) patterns were collected on a Rigaku powder X-ray diffractometer using Cu K$_\alpha$ ($\lambda$ = 1.541 Å) radiation. In order to quantitatively calculate the graphitic degree of carbon membranes prepared with different temperatures, we grinded the carbon membranes into fine powders for the XRD test, avoiding any shift of the (002) peak resulted from the unevenness of carbon membranes. X-ray photoelectron spectroscopy (XPS) data were collected by an Axis Ultra instrument (Kratos Analytical) under ultrahigh vacuum (<10$^{-8}$ Torr) and by using a monochromatic Al K$_\alpha$ X-ray source. The adventitious carbon 1s peak was calibrated at 285 eV and used as an internal standard to compensate for any charging effects. Raman measurements were performed on a Renishaw inVia Reflex with an excitation wavelength of 473 nm and laser power of 100 mW at room temperature. Nitrogen sorption isotherms were measured at -196 °C using a Micromeritics ASAP 2020M and 3020M system. Samples were degassed for 6 h at 200 °C before the measurements. Pore size distribution was calculated using the density functional theory (DFT) method. Gel permeation chromatography (GPC) was conducted at 25 °C in a NOVEMA-column with mixture of 80% acetate buffer and 20% methanol as eluent (Flow rate: 1.00 mL/min, PEO standards using RI detector-Optilab-DSP-Interferometric Refractometer). Thermal gravimetric analyses (TGA) were performed on a Netzsch TG209-F1 apparatus at a heating rate of 10 °C min$^{-1}$ under N$_2$ flow. Elemental analyses were obtained from the service of Mikroanalytisches Labor Pascher (Remagen, Germany). A field emission scanning electron microscope (FESEM, FEI Quanta 600FEG) was used to acquire SEM images. Transmission electron microscope (TEM) and high-resolution TEM (HRTEM) images, selected-area electron diffraction (SAED) patterns, and the HAADF-STEM-EDS data were taken on a JEOL JEM-2100F transmission electron microscope operated at 200 kV.

**Data availability.** The authors declare that the data supporting the findings of this study are available within the article and its Supplementary Information File.

**Acknowledgements**

Authors thank the King Abdullah University of Science and Technology (KAUST) for financial support. J. Y. is grateful for financial support from the Max Planck society, Germany, and the ERC (European Research Council) Starting Grant (project number 639720-NAPOLI).

**Author contributions**

H. W. and J. Y conceived and designed the research project. H. W. performed the experiments. T. W. supervised the research. S. M., C. M., Z. L., W. Z., Y. L., S. T., Y. H., H. Z., E. A.-H., M. N. H., J. P., W. Y., K.-W. H., L.-J. L. assisted with the characterizations of the porous carbon membranes. M. A. advised on the synthesis and electrocatalytic measurements. H. W., J. Y. and T. W co-wrote the manuscript. All authors discussed the results and commented on the manuscript.

Supplementary information is available in the online version of the paper. Reprints and permissions information is available online at www.nature.com/reprints. Correspondence and requests for materials should be addressed to J. Y. or T. W.

**Competing financial interests**

The authors declare no competing financial interests.

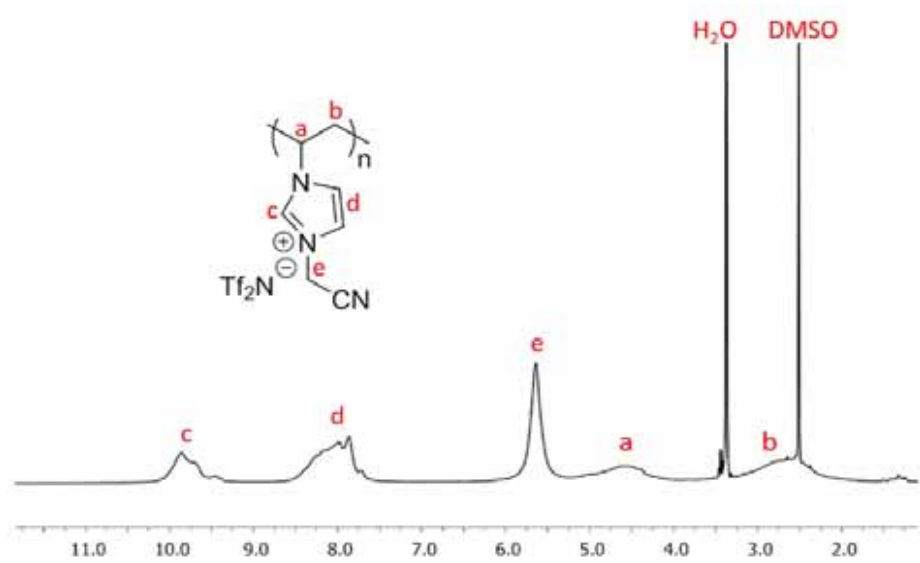

**Supplementary Figure 1 | ¹H-NMR spectra of PCMVImTf₂N in DMSO-*d₆*.**

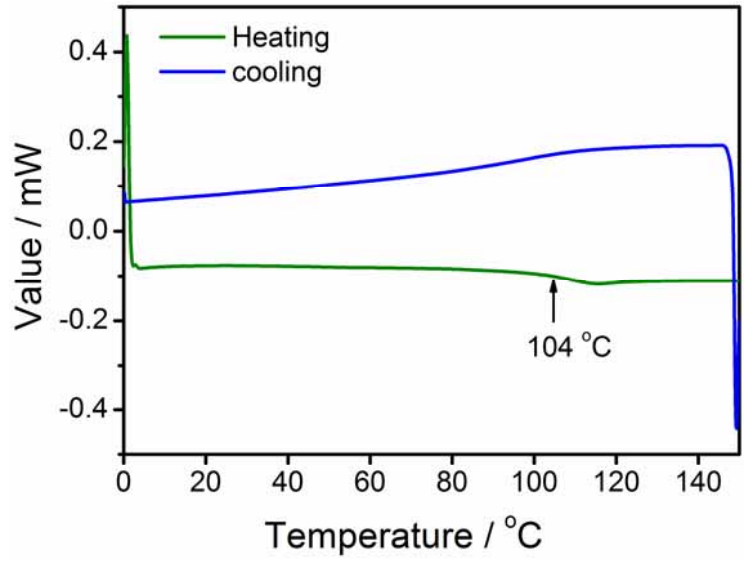

**Supplementary Figure 2 | DSC curve of PCMVImTf$_2$N, showing a glass transition temperature at about 104 $^o$C.**

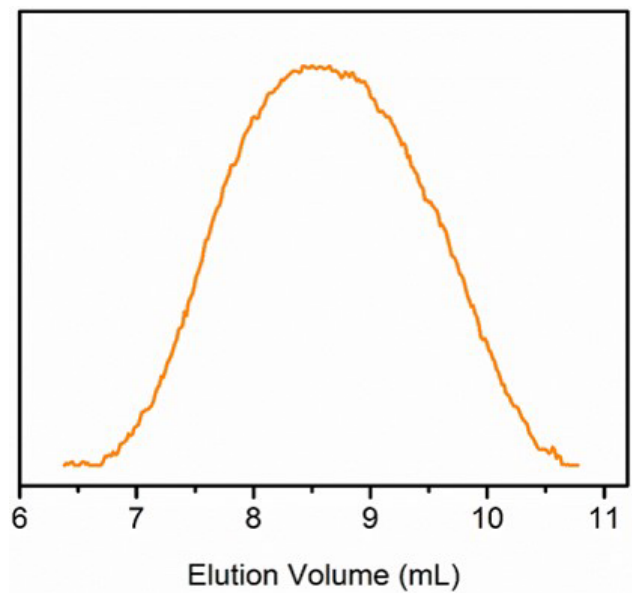

**Supplementary Figure 3 | GPC trace measured for PCMVImBr.** The apparent number-average molecular weight and PDI value of poly(3-cyanomethyl-1-vinylimidazolium bromide) (PCMVImBr) was measured to be $5.80 \times 10^5$ g/mol and 3.85, respectively (measured by GPC, eluent: water with a mixture of 80% acetate buffer and 20% methanol). Poly[3-cyanomethyl-1-vinylimidazolium bis(trifluoromethane sulfonyl)imide] (PCMVImTf$_2$N) was prepared by anion exchange of PCMVImBr with LiTf$_2$N salt. Therefore, the apparent number-average molecular weight of PCMVImTf$_2$N is calculated to be $1.12 \times 10^6$ g/mol.

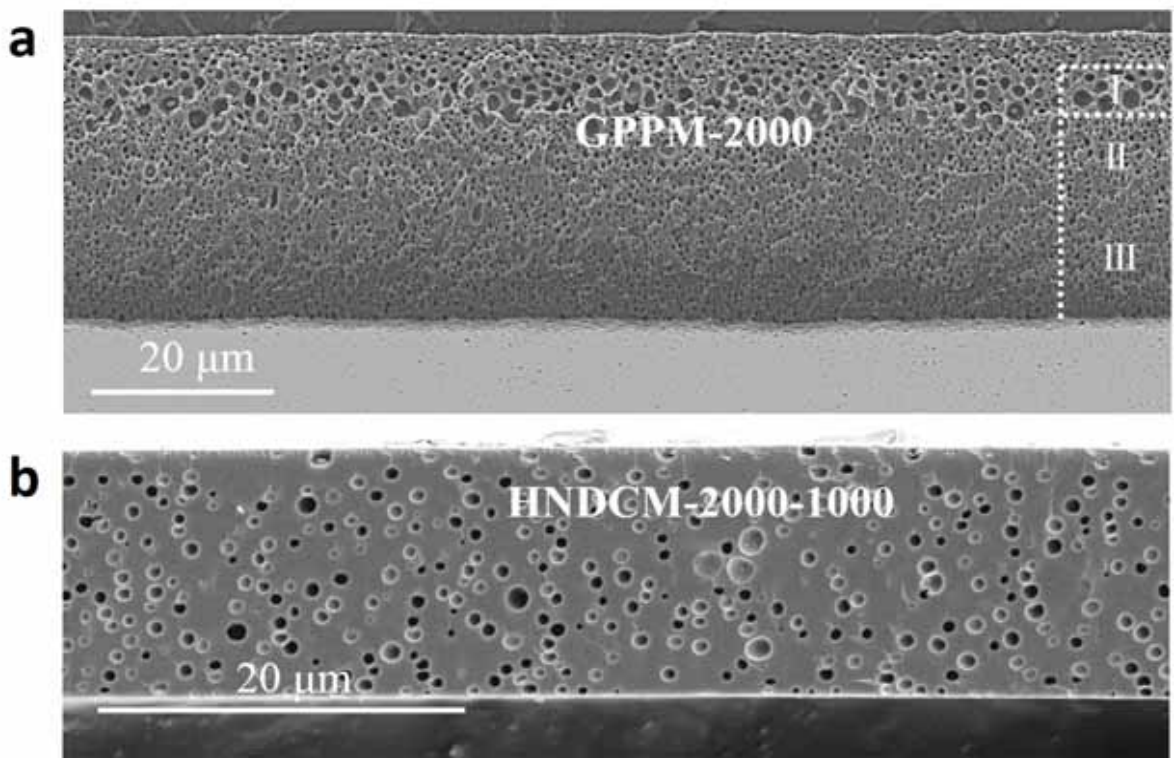

**Supplementary Figure 4 | Cross-section SEM images of GPPM-2000 and HNDCM-2000-1000. a**, Cross-section SEM image of GPPM-2000 prepared with PCMVImTf$_2$N and PAA of Mw ~ 2000 g/mol. Here, notations of GPPM-x and HNDCM-x-y are used, where x and y denote the Mw of PAA and the carbonization temperature, respectively. **b**, Cross-section SEM image of HNDCM-2000-1000. It clearly shows the pores in GPPM-2000 are continuous in a gradient distribution **(Supplementary Figure 4a)**. The average pore sizes are 2.1 μm in Zone (I), 650 nm in Zone (II), and 600 nm at the bottom, Zone III. From **Supplementary Figure 4b**, it can be clearly seen that the pores in HNDCM-2000-1000 are random and inconsecutive, indicating the morphology-maintaining carbonization can't be achieved by pyrolysis of GPPM-2000.

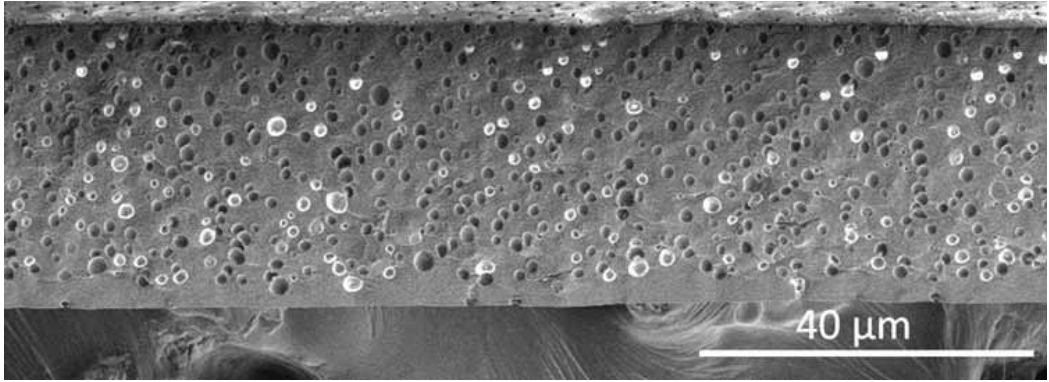

**Supplementary Figure 5 | Cross-section SEM image of the carbon sample HNDCM-2000-300 prepared from GPPM-2000 carbonized at 300 °C.**

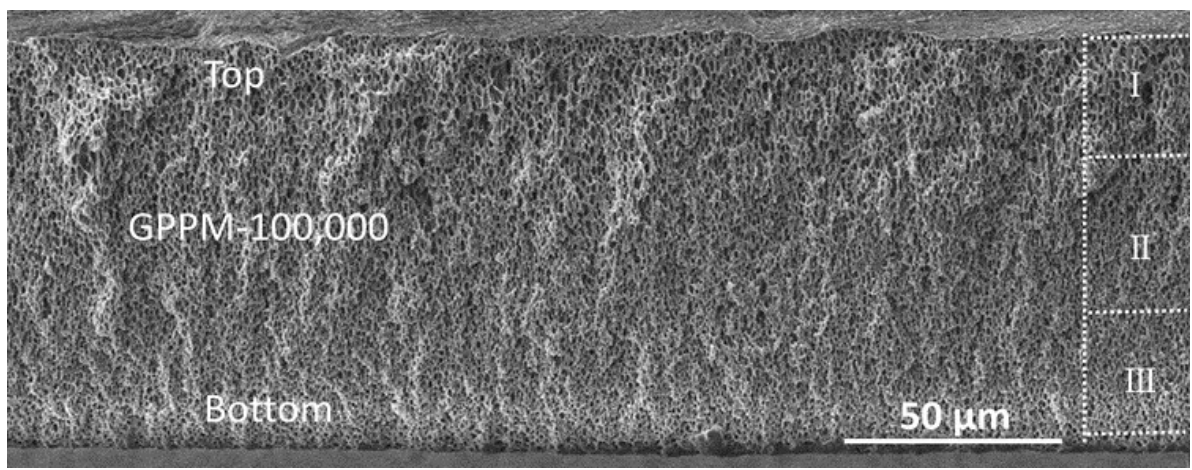

**Supplementary Figure 6 | Cross-section SEM image of the GPPM-100,000.** The pore sizes gradually decrease from the top layer (zone I, average pore size: 900 nm), to middle layer (zone II, average pore size: 740 nm) and further to bottom layer (zone , average pore ore size: 500 nm).

The formation mechanism of the gradient, hierarchically porous polymer membrane can be explained from a diffusion-controlled kinetic point of view, that is, the diffusion of aqueous $NH_3$ into the PCMVImTf$_2$N/PAA blend film from the top to the bottom is a crucial step. When the dried PCMVImTf$_2$N/PAA blend film sticking to a glass plate is immersed in aqueous $NH_3$ solution, rapid and thorough electrostatic complexation takes place in the surface region because of the direct and full contact with the $NH_3$ solution. After the first stage of full-contact electrostatic complexation, aqueous $NH_3$ gradually diffuses into the bulk membrane, neutralizes PAA and introduces interpolyelectrolyte complexation. Thus, this diffusion creates a gradient in the degree of electrostatic complexation (DEC) and correspondingly in the pore size distribution. The degree of electrostatic complexation (DEC, defined as the molar fraction of imidazolium units that undergo complexation) of GPPMs prepared with different $M_w$ of PAA are listed in **Supplementary Table 1**.

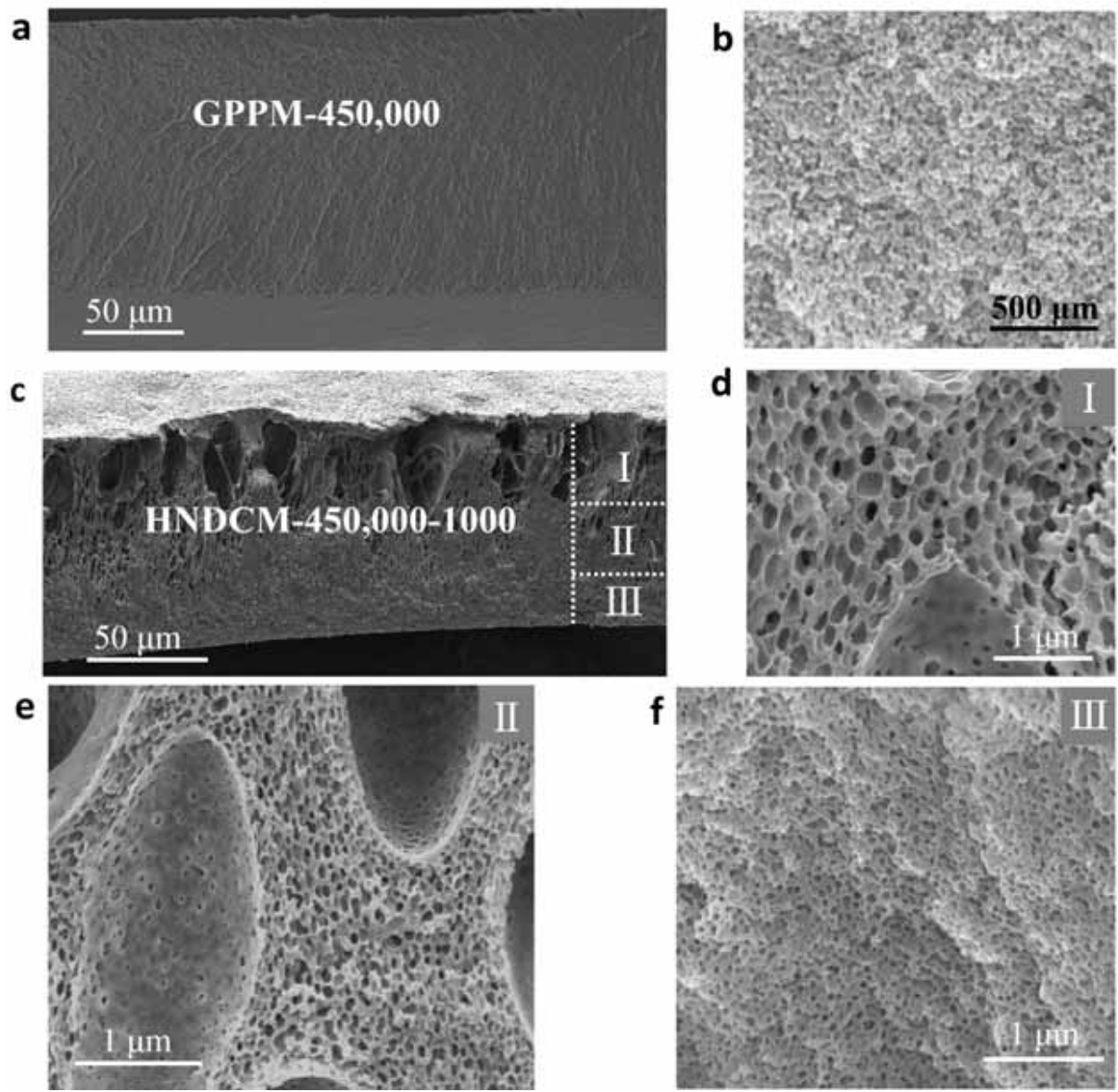

**Supplementary Figure 7 | Cross-section SEM images of GPPM-450,000 and HNDCM-450,000-1000. a**, Cross-section SEM image of GPPM-450,000; **b**, Enlarged SEM image of GPPM-450,000; **c**, Cross-Section SEM image of HNDCM-450,000-1000; **d-f**, Representative SEM images of the cross-section structures in Zone I, II and III of HNDCM-450,000-1000.

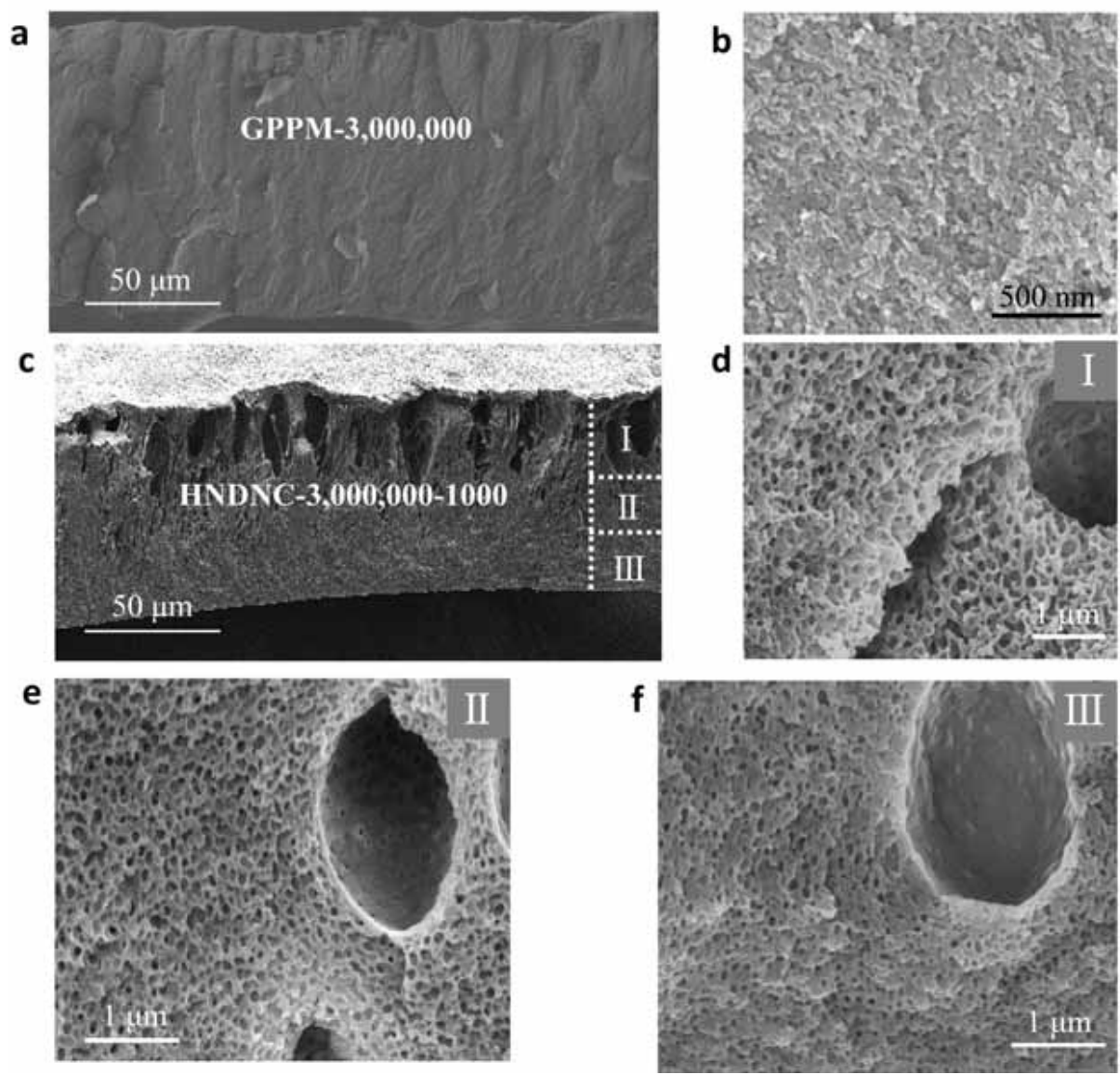

**Supplementary Figure 8 | Cross-section SEM images of GPPM-3,000,000 and HNDCM-3,000,000-1000. a**, Cross-section SEM image of GPPM-3,000,000; **b**, Enlarged SEM image of GPPM-3,000,000. **c,** Cross-section SEM image of HNDCM-3,000,000-1000; **d-f,** Representative SEM images of the cross-section structures in Zone I, II and III of HNDCM-3,000,000-1000.

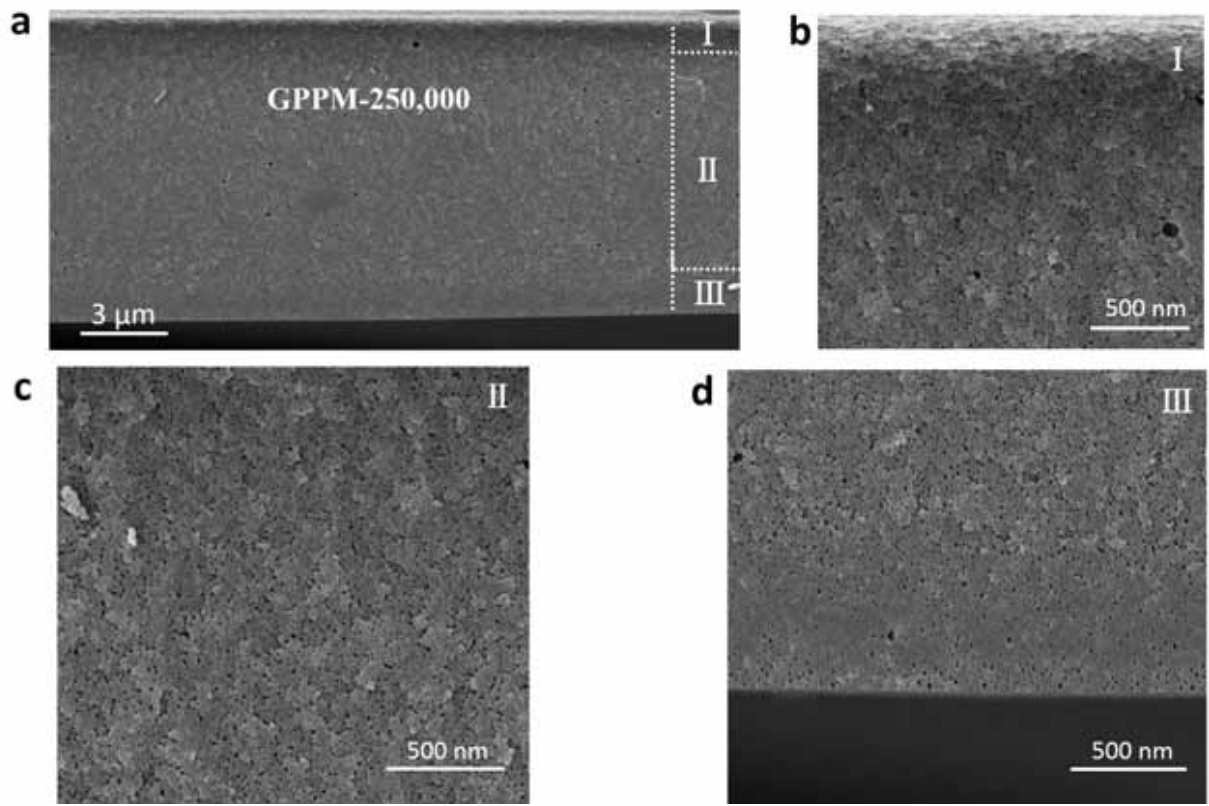

**Supplementary Figure 9 | Cross-section SEM images of GPPM-250,000 a**, Low magnification SEM image of GPPM-250,000. **b-d**, Representative SEM images of the cross-section structures in Zone I, II and III of GPPM-250,000.

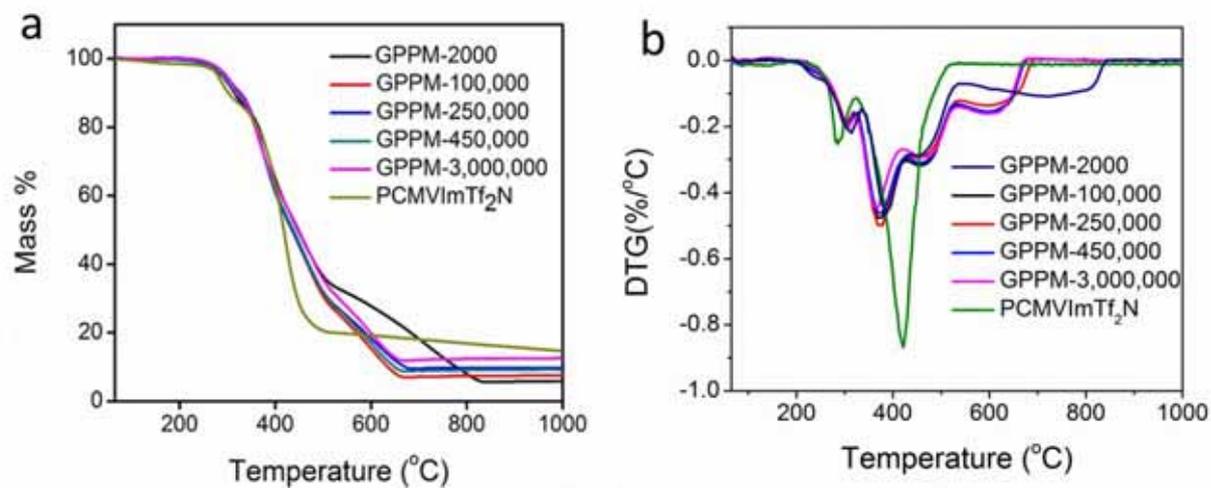

**Supplementary Figure 10 |. Thermal analysis of PCMVImTf$_2$N and GPPMs prepared with PCMVImTf$_2$N and PAA of different MWs. a,** TGA and **b**, DTA curves.

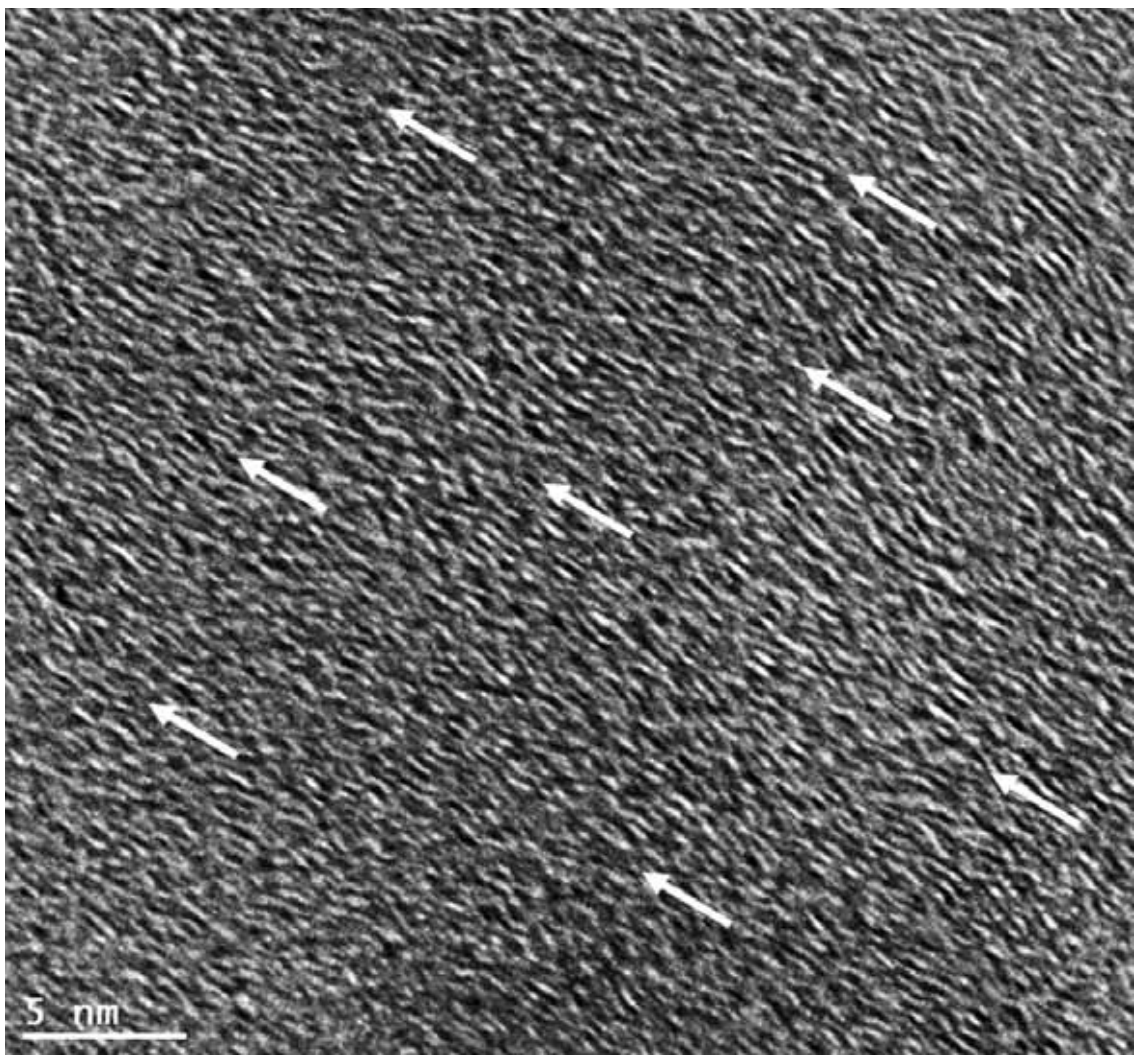

**Supplementary Figure 11 | HRTEM image of the HNDCM-100,000-800.** The white arrows point out the preferential orientation of the graphitic layers.

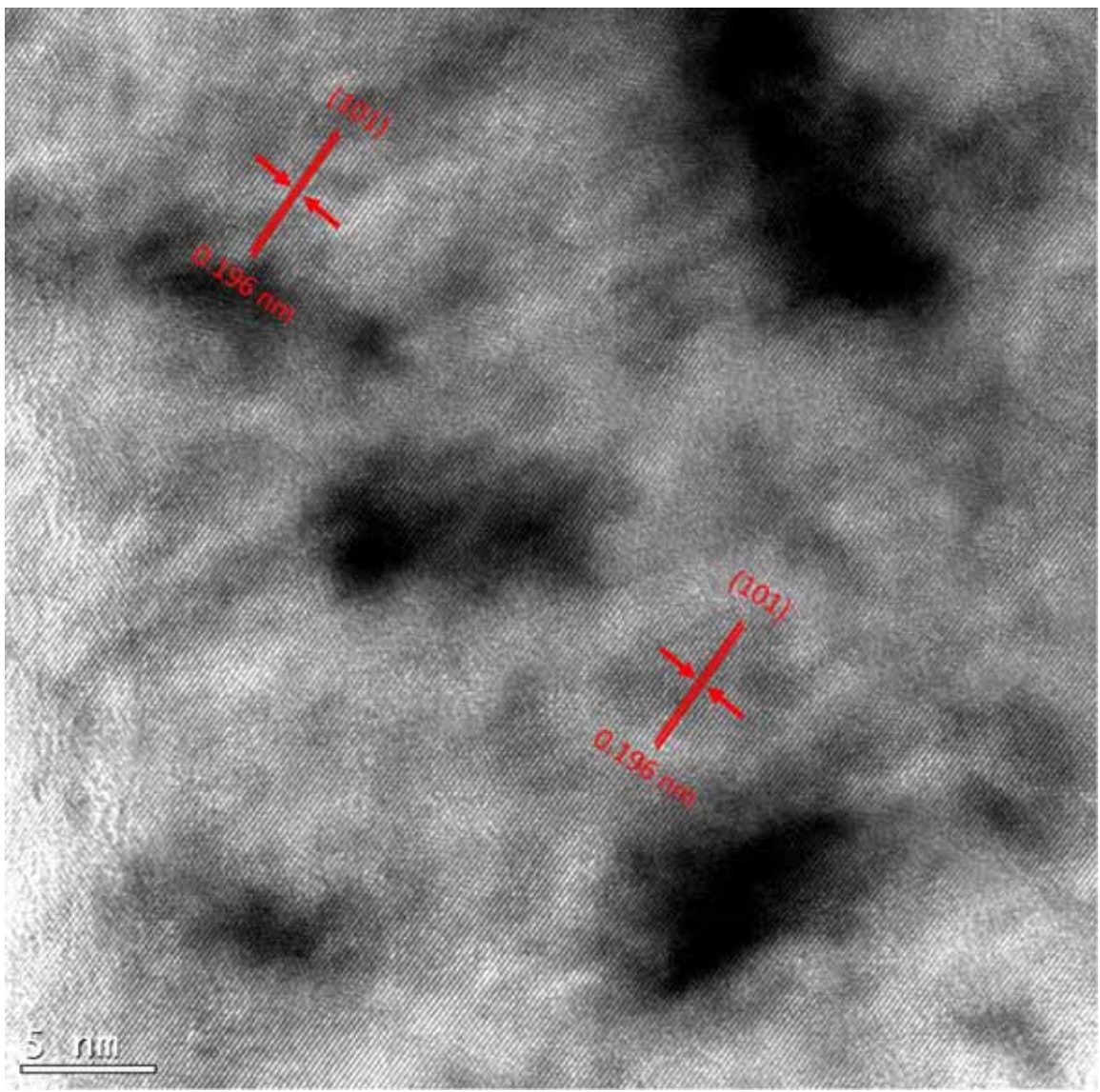

**Supplementary Figure 12 | HRTEM image of the HNDCM-100,000-900 with (101) plane dominated sheets.** It should be noted that in HNDCM-100,000-900, we observed two phases: one is the (101) plane dominated graphitic sheets in the membrane matrix, as shown here; the second phase is the (002) plane dominated concentric onion-like graphitic structures (HRTEM **Supplementary Figure 21**).

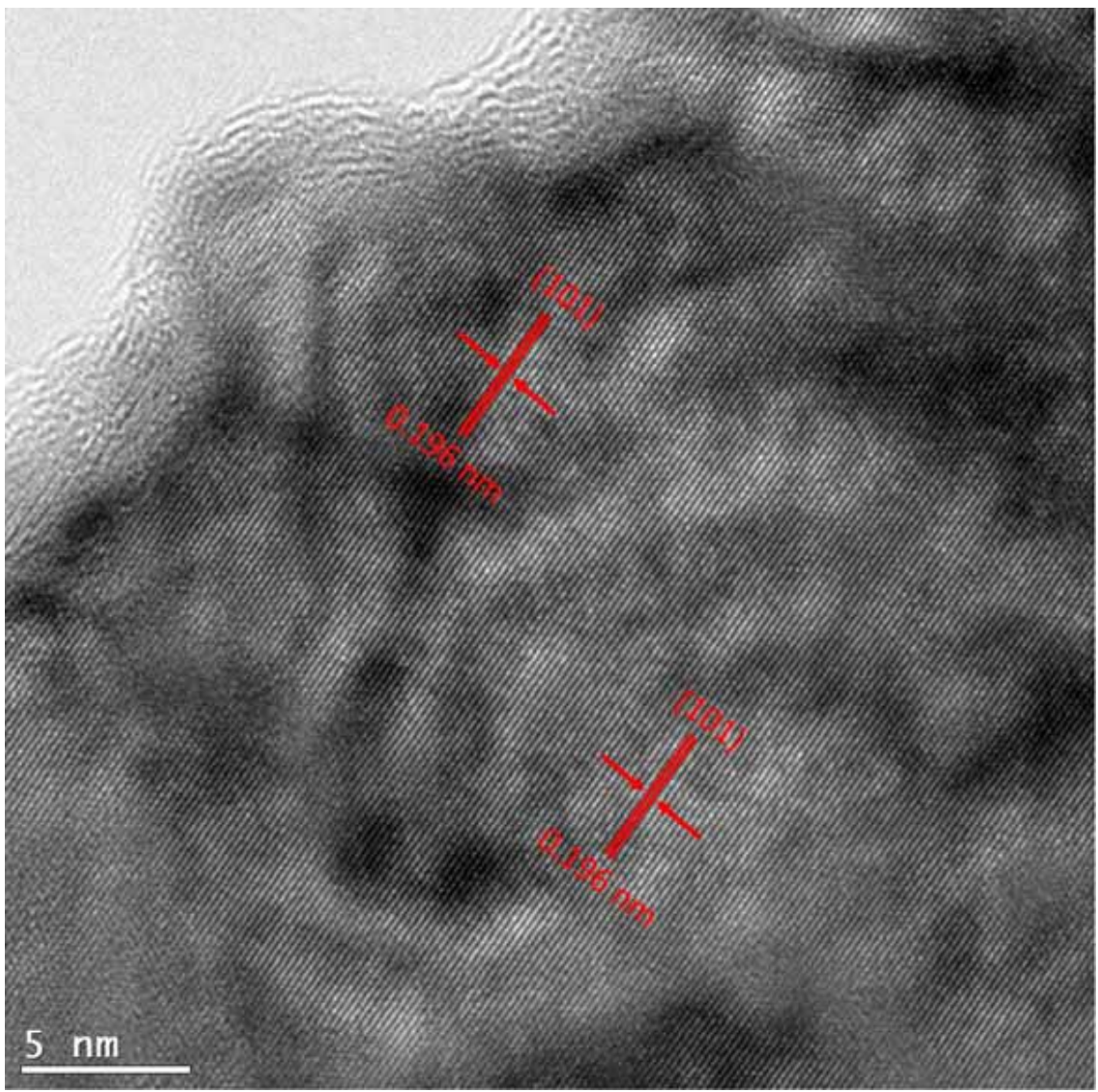

**Supplementary Figure 13 | HRTEM image of the HNDCM-100,000-1000 with (101) plane dominated sheets.** Similar to HNDCM-100,000-900, in HNDCM-100,000-1000, we also observed two phases: one is the (101) plane dominated graphitic sheets, as shown here; the second phase is the (002) plane dominated concentric onion-like graphitic structures (HRTEM **Supplementary Figure 22**).

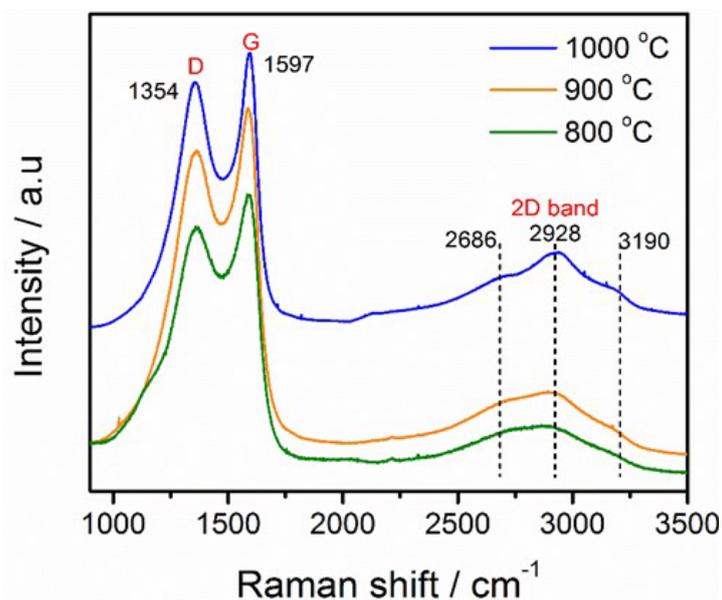

**Supplementary Figure 14| Raman spectra of HNDCM-100,000-y (y=800, 900, and 1000).** The Raman spectra of the HNDCM-100,000-y samples (y=800, 900, 1000) contain two bands at 1354 and 1597 cm$^{-1}$, which were assigned to the typical disorder band (D band) and graphitic band (G band) of carbon, respectively. The $I_D/I_G$ ratio for all three samples was ~0.85, indicating their structural similarity[1]. The 2D band is Raman active for crystalline graphitic carbons and sensitive to the π band in the graphitic electronic structure[2]. The 2D peak became much sharper as the carbonization temperature increased, and the most intense peak was observed for HNDCM-100,000-1000.

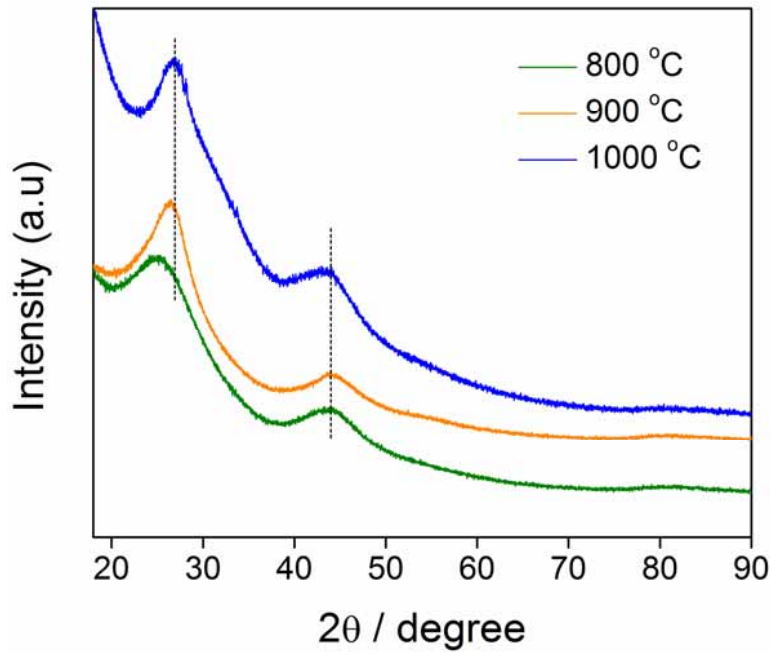

**Supplementary Figure 15 | XRD patterns of HNDCM-100,000-y (y=800, 900, and 1000)**

We observed the crystalline phase in HNDCM-100,000-900/1000 samples in TEM, which can only reflect local structure information. To obtain a general picture of the crystalline phases of HNDCM-100,000-900/1000, the two samples were further analyzed by X-ray diffraction[3]. The graphitization index can be used to characterize quantitatively the degree of graphitization of carbon materials. The graphitization index is derived from the average interplanar spacing between two successive graphite layers according to the equation 1:

$$g_p = \frac{3.440 - d_{002}}{3.440 - 3.354} \quad \cdots\cdots\cdots\cdots \text{Equation 1}$$

According to the Bragg's law, the interplanar spacing is given by $d_{002} = \lambda/\sin\theta$, where $\lambda$ is the wavelength of the incident X-ray beam. The copper $K_\alpha$ line is 1.541 Å. The $\theta$ values of HNDCM-100,000-900 and HNDCM-100,000-1000 are 26.74° and 26.93°, respectively. Correspondingly, the $d_{002}$ values of HNDCM-100,000-900 and HNDCM-100,000-1000 are 3.42 and 3.40 Å, respectively. The graphitization index of HNDCM-100,000-900 and HNDCM-100,000-1000 are calculated to be 0.23 and 0.46, respectively. This indicates the degree of crystallinity of carbon membranes is favored expectedly by higher pyrolysis temperature. It should be mentioned that due to the incorporation of nitrogen atoms into the graphitic phase, the interplaner spacing in nitrogen doped carbons is enlarged to accommodate the lone electron pair on the nitrogen atoms.

In addition to the position, the width of the (002) band provides useful information to estimate the average grain size of graphitic phases. On the basis of the width, the coherence lengths Lc and La can be estimated by the Debye-Sherrer equations:

$$Lc = k\lambda / \beta \cos\theta \qquad \text{Equation 2}$$

$$La = 1.84\, k\lambda / \beta \cos\theta \qquad \text{Equation 3}$$

where k is the shape coefficient, usually k is set to 1. $\lambda$ is the wavelength of the incident beam, $\theta$ is the Bragg angle, and $\beta$ is the full width at the half maximum (fwhm). The calculated Lc/La values for the HNDCM-100,000-900 and HNDCM-100,000-1000 are 0.44 nm/0.81nm and 0.58 nm/1.06 nm, respectively.

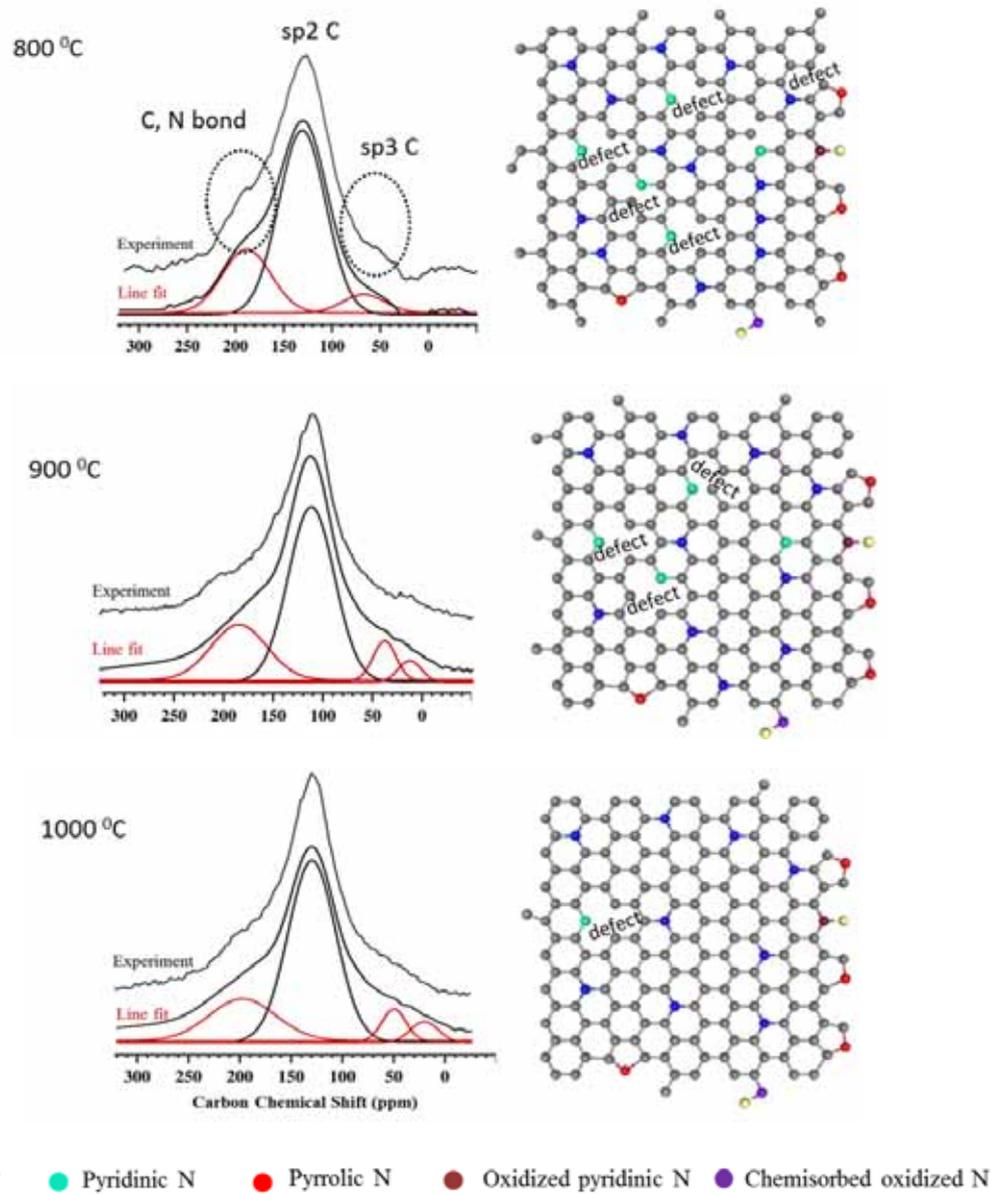

**Supplementary Figure 16 | Solid-state $^{13}$C-NMR of HNDCM-100,000-y (y=800, 900, and 1000).** Solid state $^{13}$C-NMR spectroscopies afford the qualitative and quantitative analysis of HNDCM structures. The fitted lines show that the content of carbon bonded nitrogen and sp$^3$ hybrid C decreases while the content of sp$^2$ hybrid C increases with increasing carbonization temperature from 800 to 1000 °C, indicating that higher carbonization temperature can result in higher degree of graphitization. The result that the content of carbon bonded nitrogen decreases with increasing temperature is in agreement with the elemental analysis.

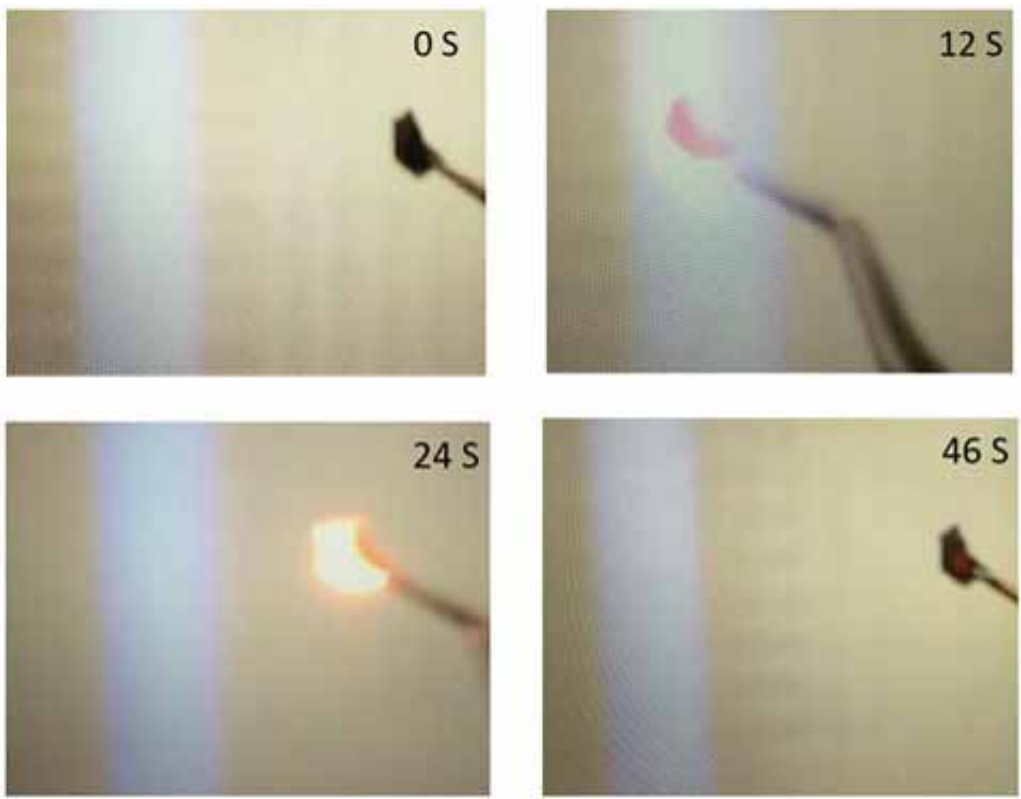

**Supplementary Figure 17 | Fire retardant property of HNDCM-100,000-1000.** Illustration of the fire retardant property of the nitrogen doped porous carbon membrane by firing the carbon membrane with an acetylene gas burner (flame temperature above 1000 °C) for 24 seconds. The membrane was found to turn light red in the flame but resumed its native black color 20s after being pulled back to air.

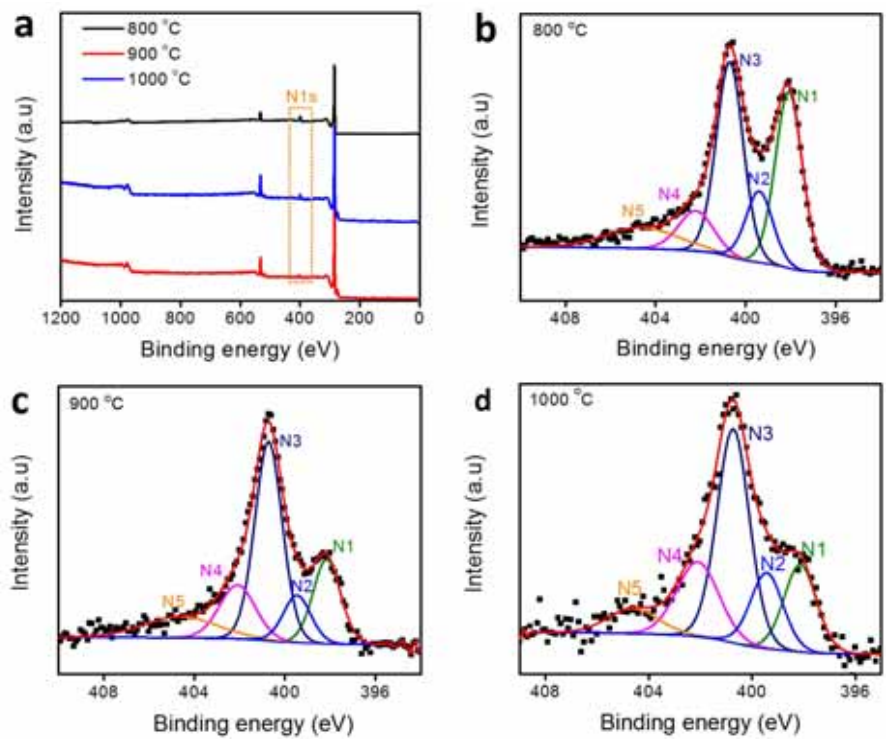

**Supplementary Figure 18 | XPS characterization of HNDCM-100,000-y (y=800, 900, and 1000) samples. a**, XPS spectra of HNDCM-100,000-y (y=800, 900, and 1000). **b-d**, The fitted XPS peaks for N1s orbit of HNDCM-100,000-y (y=800, 900, and 1000).

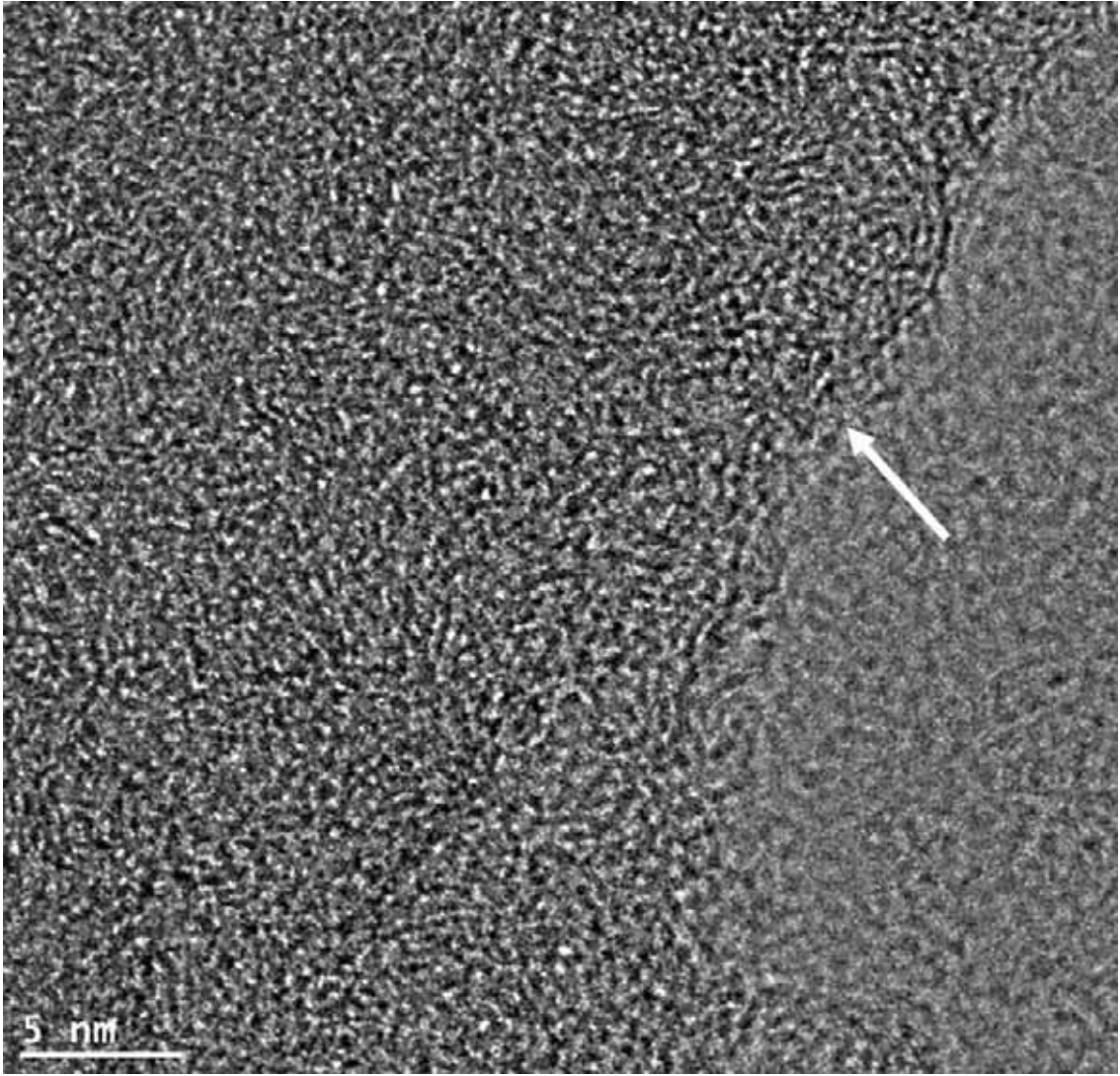

**Supplementary Figure 19 | HRTEM of carbon prepared by direct pyrolysis of native nonporous PCMVImTf₂N at 1000 °C**. The white arrow indicates the N-doped carbon.

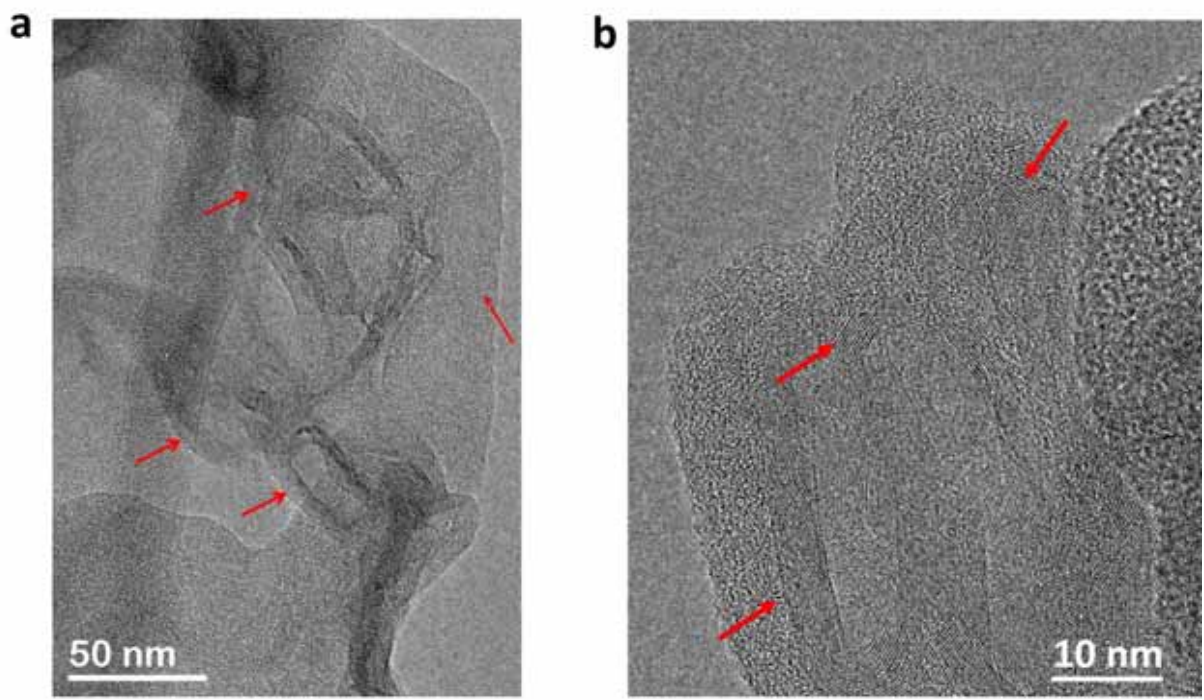

**Supplementary Figure 20 | TEM images of the HNDCM-2000-1000. a**, Low-magnification TEM image of the HNDCM-2000-1000; **b**, HRTEM image of HNDCM-2000-1000. Red arrows indicate the graphitic N-doped carbon.

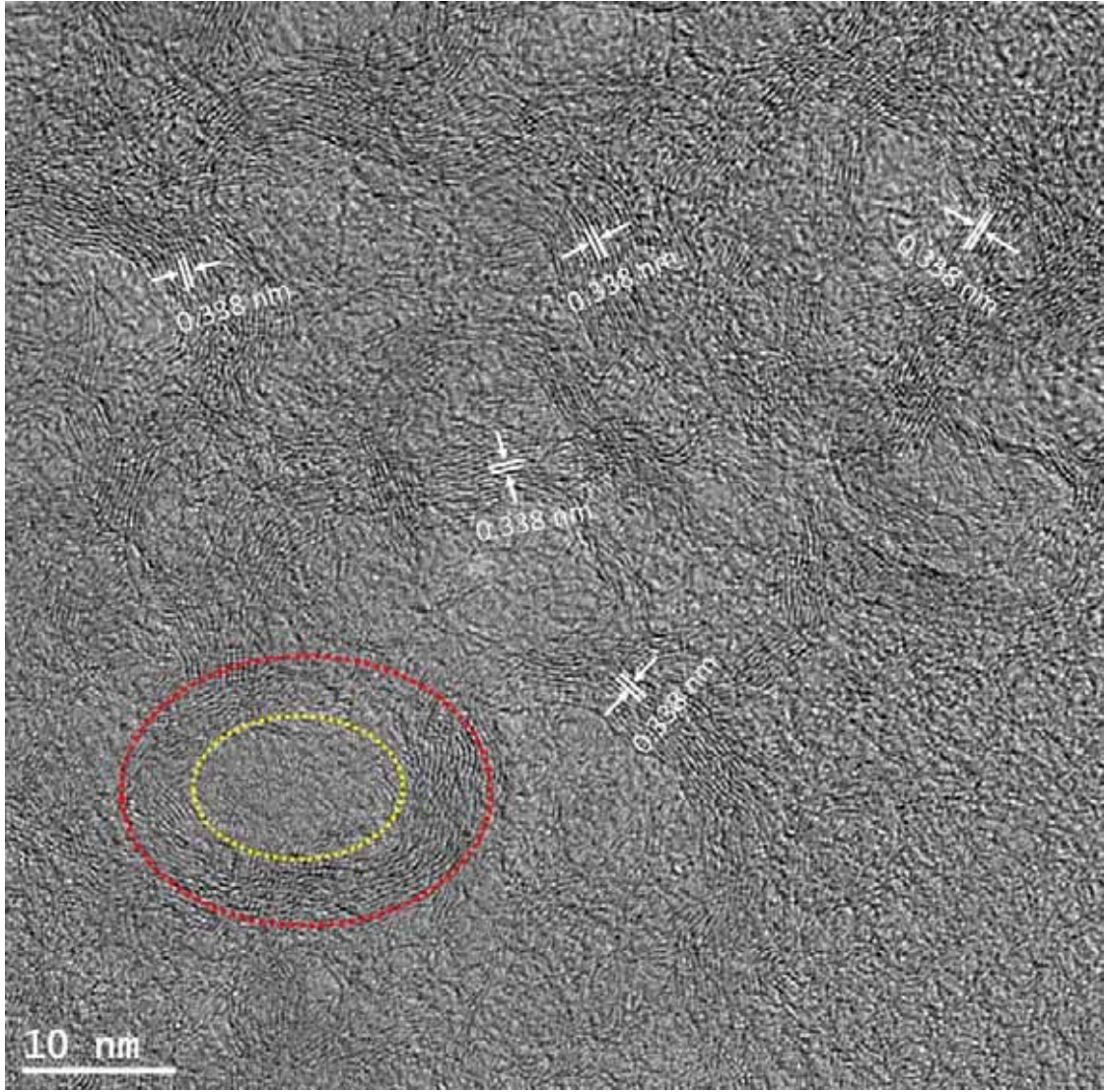

**Supplementary Figure 21 | HRTEM image of HNDCM-100,000-900 with (002) plane dominated concentric onion-like graphitic domains.** Concentric onion-like graphitic nanostructures with multi-shells (red line) and hollow cages (yellow line) are observable in the HNDCM-100,000-900.

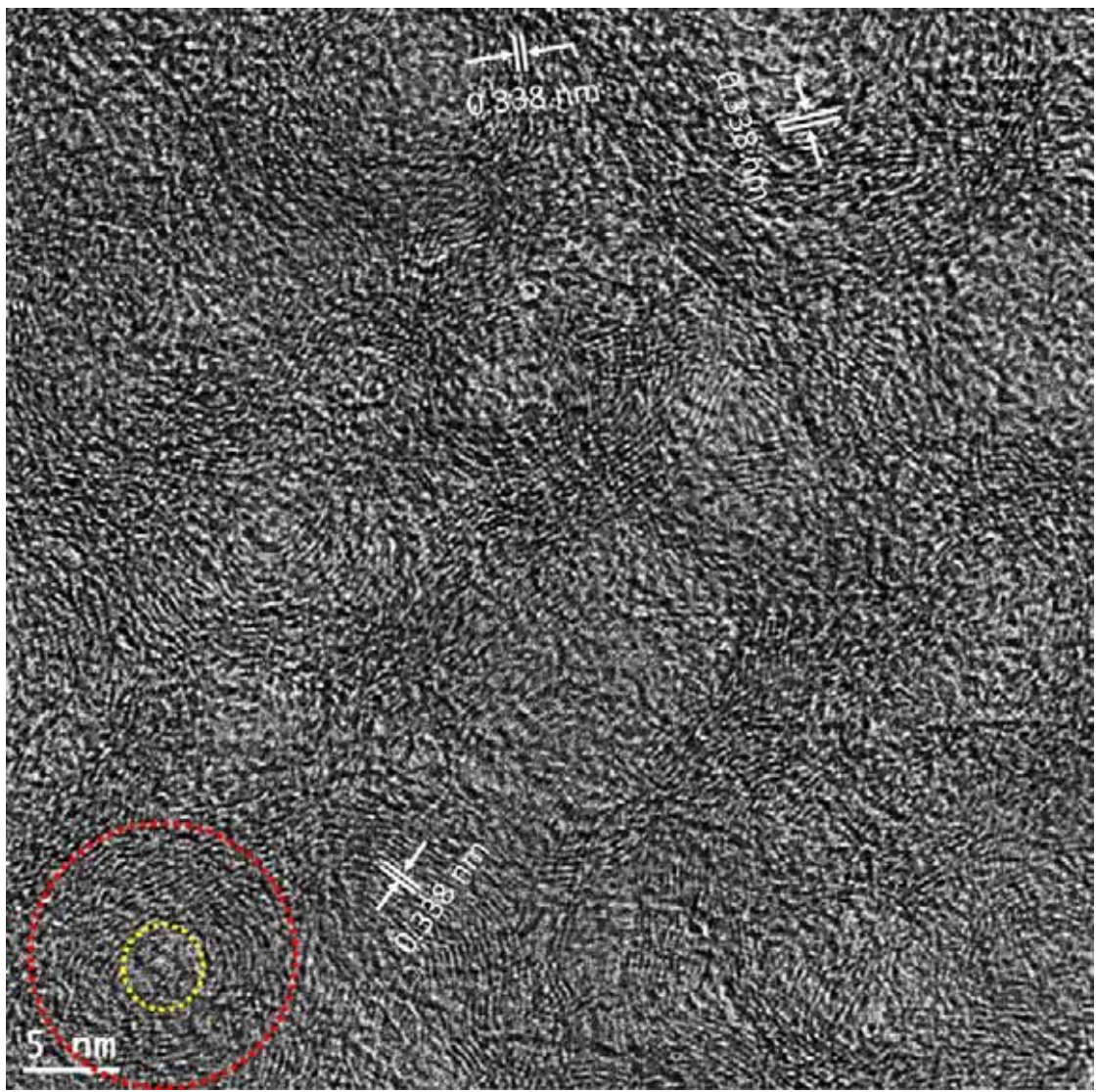

**Supplementary Figure 22 | HRTEM image of HNDCM-100,000-1000 with (002) plane dominated concentric onion-like graphitic nanostructures.** It can be seen that the typical concentric onion-like graphitic nanostructures with multi-shells (red line) and hollow cages (yellow line) exist in HNDCM-100,000-1000.

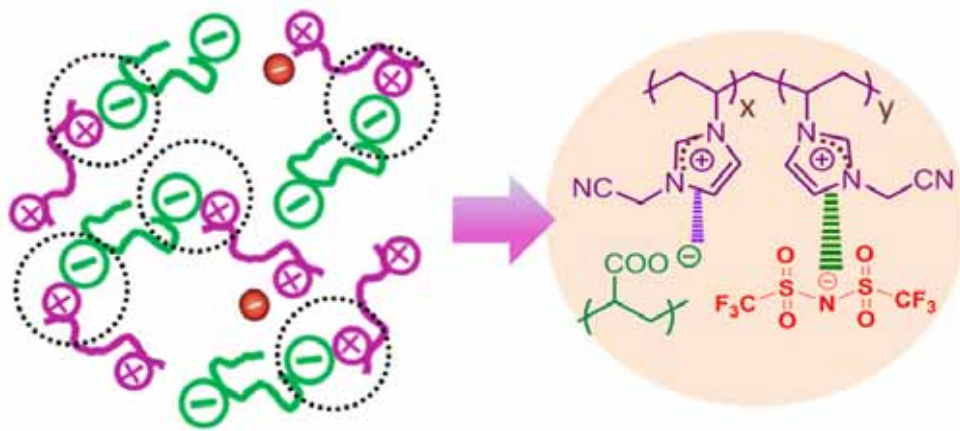

DEC= X/(X+Y)  ·················· Equation 4

DEC= (482S-64)/(287S-64)  ·················· Equation 5

**Supplementary Figure 23 | Scheme to illustrate the definition of the degree of electrostatic complexation (DEC) of the GPPM-100,000-1000.**

Equations for defining and calculating DEC. In equation S4, X denotes the imidazolium units that undergo electrostatic complexation with COO$^-$ groups on PAA; Y denotes the imidazolium units in the membrane that are not involved in the electrostatic complexation. In equation S5, S denotes the sulfur weight content. The results of the elemental analyst show that the S content is 12% in GPPM-100,000. We can calculate the DEC of GPPM-100,000-1000 is 20.8%. The content of the bis(trifluoromethanesulfonyl)imide (Tf$_2$N$^-$) can be calculated as 28.3 mol % and 53.8% wt% in the GPPM-100,000.

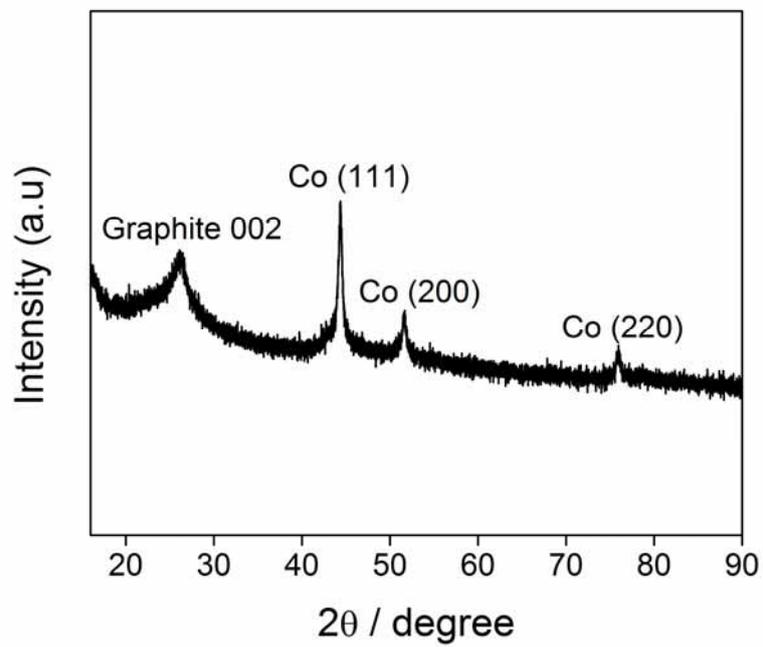

**Supplementary Figure 24 | XRD patterns of HNDCM-100,000-1000/Co.**

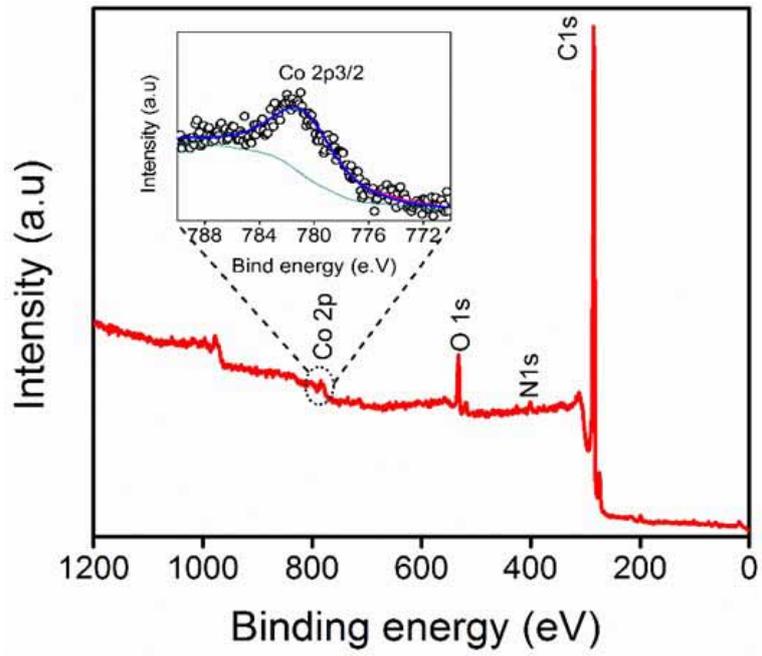

**Supplementary Figure 25 | XPS spectra of HNDCM-100,000-1000/Co, inset is the Co 2p3/2 peak.**

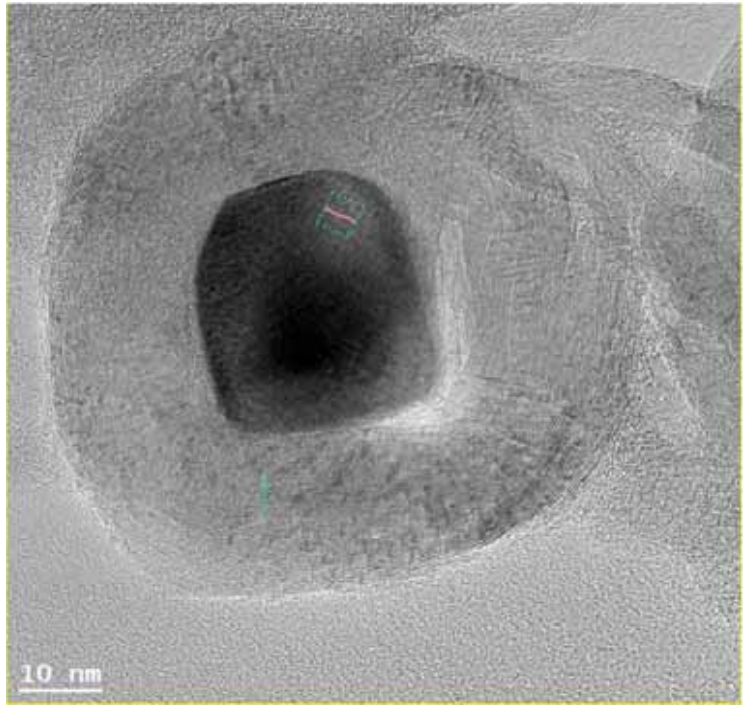
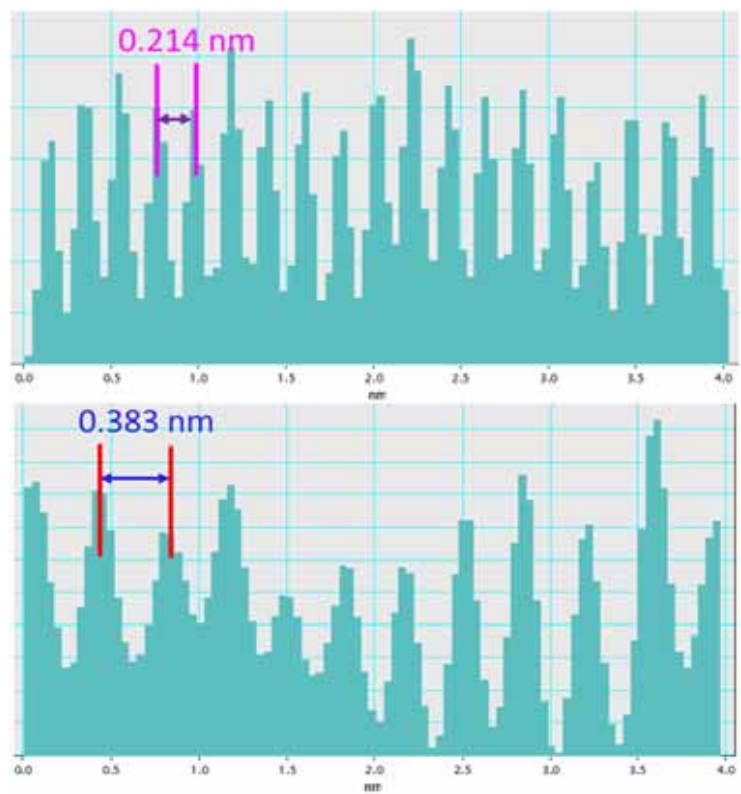

**Supplementary Figure 26 | HRTEM image of the Co nanoparticles covered by a thin graphitic carbon shell of several nm in thickness.** The lattice *d*-spacing of 0.214 nm (medium) and 0.383 nm (bottom) are corresponding to the $\{10\bar{1}10\}$ plane in hcp-Co and graphite, respectively.

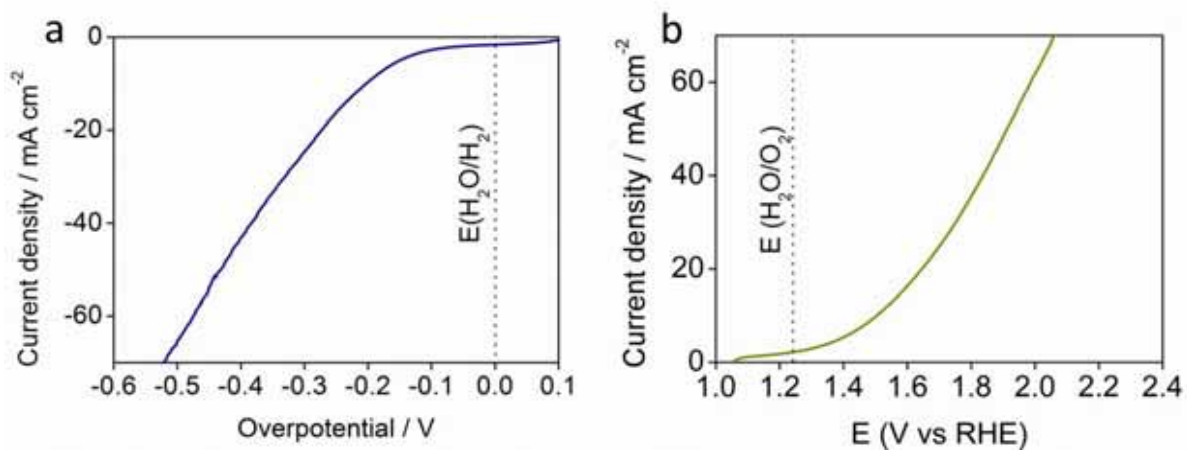

**Supplementary Figure 27 | Without IR-corrected LSV curves of (a) HER and (b) OER for sample HNDCM-100,000-1000/Co, respectively.**

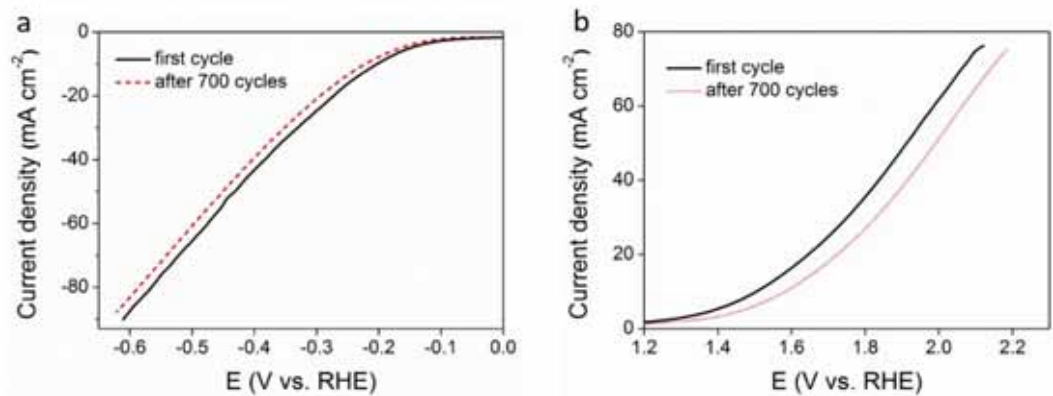

**Supplementary Figure 28 | The CV stability curves of the 100,000-1000/Co electrode in HER (a) and OER (b), respectively, before and after running the CV test for 700 cycles.**

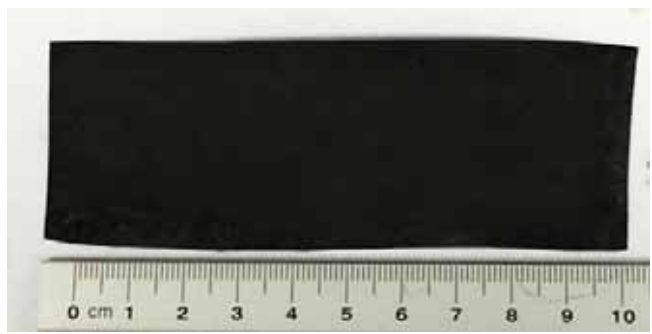

**Supplementary Figure 29 | A large piece of HNDCM-100,000-1000/Co catalyst with a size of 10.5 x 3.5 cm² and thickness of ~70 μm.** This membrane is the largest one that we can prepare using our carbonization oven at 1000 °C in our lab. The membrane size can be even larger if larger carbonization ovens are available.

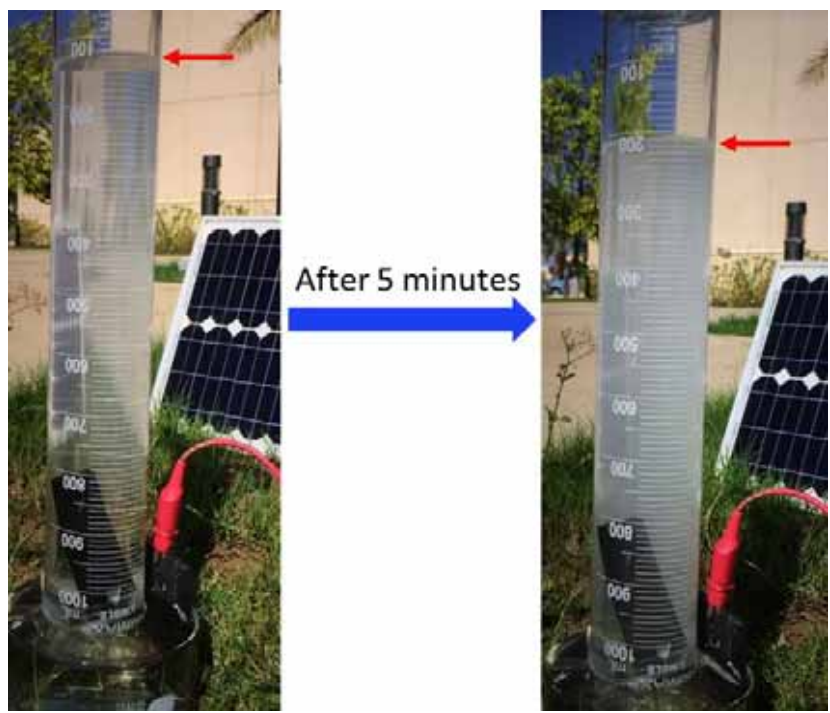

**Supplementary Figure 30 | Photographs taken before and after HER driven by solar cell in 1 M KOH;** 90 mL $H_2$ was collected within 5 minutes. The red arrows indicate the top line of the solution. (Note: this test was performed with unetched HNDCM-100,000-1000/Co film, which is even less efficient in $H_2$ production than the HCl-etched one.

**Supplementary Table 1 | Degree of electrostatic complexation (DEC) of the GPPMs calculated by equations shown in Supplementary Fig. 23.**

| GPPM | S content determined by elemental analysis | DEC in average |
|---|---|---|
| GPPM-2000 | 12.6 % | 11.7 % |
| GPPM-100,000 | 12.0 % | 20.8 % |
| GPPM-250,000 | 12.1 % | 19.4 % |
| GPPM-450,000 | 11.7 % | 24.9 % |
| GPPM-3,000,000 | 11.8 % | 23.6 % |

**Supplementary Table 2 | HER performance of HNDCM-100,000-1000/Co in this work, in comparison with several representative results with high performance non-noble metal based catalysts from recent publications.**

| Catalyst | Current density j (mA cm$^{-2}$) | Overpotential (vs. RHE) at the corresponding j | Condition | *References* |
|---|---|---|---|---|
| MoB | 10 | 225 mV | alkaline | *Angew. Chem., Int. Ed.* **51**, (12703-12706) 2012. (S4) |
| MoC | 10 | > 250 mV | alkaline | *Angew. Chem. Int. Ed.* 126, (6525–6528), 2014. (S5) |
| **Co-NRCNT** | **10** | **370 mV** | **alkaline** | ***Angew. Chem., Int. Ed.*, 53, (4372-) 2014. (S6)** |
| HNDCM-100,000-1000/Co | 10 | 158 mV | alkaline | *This work* |
| CoOx@CN | 10 | 232 mV | alkaline | *J. Am. Chem. Soc.* **137**, (2688−2694) 2015 (S7) |
| Nanoporous MoS$_2$ | 10 | 270 mV | acid | *Nature Mater.* **11**, (963-969) 2012. (S8) |
| Au supported MoS$_2$ | 0.2 | 150 mV | acid | *Science* **317**, (100-102) 2007. (S9) |
| Exfoliated WS$_2$/MoS$_2$ nanosheets | 10 | 187-210 mV | acid | *Nature Mater.* **12**, (850-855) 2013. (S10)*; J. Am. Chem. Soc.* **135**, (10274-10277) 2013. (S11) |
| MnNi | 10 | 360 mV | Alkaline | *Adv. Funct. Mater.* **25**, (393-399) 2015. (S12) |

**Supplementary Table 3 | Data collected for Supplementary Fig. 18, the normalized results of different N contents**

| Samples | N1 % | N 2% | N 3% | N4 % | N 5% |
|---|---|---|---|---|---|
| HNDCM-100,000-800 | 32.1 | 13.0 | 35.3 | 9.4 | 10.2 |
| HNDCM-100,000-900 | 18.5 | 10.4 | 43.6 | 15.3 | 12.2 |
| HNDCM-100,000-1000 | 17.1 | 14.8 | 41.6 | 18.9 | 7.6 |

The fitted XPS peaks for N1s orbit of HNDCM-100,000-y (y=800, 900, and 1000) can be deconvoluted into five different bands at ~398.1, 399.5, 400.7, 402.1, and 404.6 eV, which correspond to pyridinic (N1), pyrrolic (N2), graphitic (N3), oxidized pyridinic (N4) and chemisorbed oxidized nitrogen (N5), respectively. These various N species lead to different chemical/electronic environments of neighboring carbon atoms and hence different electro-catalytic activities. The curve fitting and the corresponding normalized results indicate a conversion from pyridinic to graphitic nitrogen with increasing temperature, for example, the contents of pyridinic N in HNDCM-100,000-800, HNDCM-100,000-900 and HNDCM-100,000-1000 are 32.1%, 18.5% and 17.1%, respectively, which is consistent with previous reports on N-doped carbon materials[13].